\documentclass[preprint]{iucr} 
  \journalcode{M}

                \def\href#1{\relax}
\ifPDF
\RequirePackage{hyperref}
\PassOptionsToPackage{pdftex,bookmarksopen,bookmarksnumbered}{hyperref}
\fi

\RequirePackage{graphicx}
\usepackage{graphicx}
\usepackage{amsfonts}
\usepackage{amssymb}
\usepackage{amsmath}
\usepackage{color}
\usepackage{upgreek}
\usepackage{tabularx}
\usepackage{bm}
\usepackage{amstext}
\usepackage{amsfonts}
\usepackage{dcolumn} 

\begin{document}                  



\title{Atomic structure and formation mechanism of a newly discovered charge density wave in the $m$=2 monophosphate tungsten bronze}
\shorttitle{CDW in $P_4W_4O_{20}$}

     
\cauthor[a,b,c]{Arianna}{Minelli}{minellia@ornl.gov}{}
\author[d,e]{Elen}{ Duverger-Nedellec}
\aff[a]{European Synchrotron Radiation Facility, BP 220, F-38043 Grenoble Cedex, \country{France}}
\aff[b]{Department of Chemistry, University of Oxford, South Parks Road, Oxford OX1 3QR, \country{U.K.}}
\aff[c]{Neutron Scattering Division, Oak Ridge National Laboratory, Oak Ridge, TN 37831, \country{USA}}
\aff[d]{Univ. Bordeaux, CNRS, Bordeaux INP, ICMCB, UMR 5026, F-33600 Pessac, \country{France}}
\aff[e]{CRISMAT-ENSICAEN, University of Caen Basse-Normandie, CNRS/UMR 6508, 6 Bd Mar\'echal Juin, 14050 CAEN Cedex 4, \country{France}}
\author[e]{Olivier}{Perez}
\author[e]{Alain}{Pautrat}
\author[f]{Adrien}{Girard}
\aff[f]{Sorbonne Universit\'e, CNRS, MONARIS, F-75252 Paris, France}
\author[a]{Johnathan}{Bulled}
\author[g]{Marek}{Mihalkovi{\v{c}}}
\aff[g]{Institute of Physics, Slovak Academy of Sciences, D\'ubravsk\`a cesta 9, Bratislava 84511, \country{Slovak Republic}}
\author[h]{Marc}{de Boissieu}
\aff[h]{University of Grenoble Alpes, CNRS, SIMAP, 38000 Grenoble, \country{France}}
\author[a]{Alexei}{Bosak}

\maketitle

\begin{abstract}
The $m$=2 member of the monophosphate tungsten bronze family has been considered the only one in the family without an electronic instability at low temperature. In this paper, we report the discovery of a charge density wave phase in this compound, with a transition temperature of 290 K and an incommensurate modulation vector \textbf{q}=0.245\textbf{b*}+ $\upxi$\textbf{c*}. The presence of this new phase is confirmed by diffraction and resistivity measurements. Pre-transitional dynamics are investigated using diffuse and inelastic x-ray scattering, revealing a clear Kohn anomaly. We analyze both structural and electronic contributions to the phase transition, providing a comprehensive picture of the mechanism driving this newly identified instability.
\end{abstract}

\section{INTRODUCTION}


Low-dimensional metals exhibit peculiar electronic phases, such as superconductivity and charge or spin density waves (C-SDW). The charge density wave (CDW) phase, where electrons condense in momentum space, was first linked to an instability described by Peierls in 1955~\cite{Peierls1955}. In his original work, Peierls introduced this intrinsic instability in a one-dimensional (1D) chain as a mathematical concept, without initially associating it with a real physical structure. In this 1D system, the parabolic electronic band of non-interacting particles - subjected to a periodic external potential - has a Fermi wavevector determined by the chemical potential. For a half-filled band, the system becomes unstable with respect to a doubling of the unit cell, which leads to the opening of an energy gap. Subsequently, this type of instability has been observed in many real materials that exhibit quasi-1D behavior and linear-chain correlations~\cite{Peierls1991}. \\

Many reviews classify charge density waves (CDWs) in various ways~\cite{Rossnagel2011, Zhu2015, Pouget2024}. These classifications are generally qualitative, as a complete microscopic theory is still lacking. However, as noted by Rossnagel~\cite{Rossnagel2011}, we should mention the microscopic theories proposed in the book edited by Motizuki~\cite{Motizuki1986}. In this paper, we adopt a commonly used notation that categorizes CDWs into two types: those with i) weak and those with ii) strong electron-phonon coupling. The weak coupling case can be well described by the Peierls model, in which the lattice distortion and CDW — $i.e.$, phonons and electrons - are only weakly coupled. In contrast, the strong coupling regime is characterized by a larger amplitude of distortion, a wider energy gap, and a different behavior of the Kohn anomaly. It is difficult to objectively classify these two types based solely on experimental observables. However, one useful parameter is the ratio between the electronic energy gap, $\Delta$, and the Debye phonon frequency, $\omega_D$, which reflects the role of the Fermi surface in driving the CDW. This contribution decreases with increasing coupling strength.\\

Typically, in studies of CDW phase transitions, structural and electronic effects are considered as arising from the same instability, rather than being treated as distinct contributions. Real structures, even in low-dimensional metals, deviate significantly from the idealized one-dimensional metallic chain. In quasi-1D systems, deviations from simple Peierls-type theory must be considered. As Johannes and Mazin, as later Pouget and Canadell, have argued~\cite{Johannes2008,Pouget2024}, it is not possible to fully describe the complexity of CDW transitions using only the Peierls mechanism and the concept of perfect Fermi surface nesting.\\

To highlight the need for a detailed analysis that separates structural and electronic contributions to the phase transition and thus, their contribution to the instability that generates the CDW phase, we studied a relatively simple oxide with a quasi-1D electronic structure. The lowest member of the monophosphate tungsten bronze family, (PO$_2$)$_4$(WO$_3$)$_{2m}$, consists of zigzag chains of WO$_6$ octahedra surrounded by PO$_4$ tetrahedra. Other members of this family show a ReO$_3$-type layered structure composed of tungsten-oxide slabs and phosphate tetrahedral monolayers, leading to a quasi-2D instability. Our combined structural, electronic, and dynamical studies on this compound provide a comprehensive picture of the phase transition, allowing us to disentangle the contributions from the lattice and electronic subsystems. Moreover, our results show that the structural instability is not entirely driven by the electronic CDW. This analysis is made possible through diffraction, diffuse scattering, physical measurements, inelastic X-ray scattering, and molecular dynamics simulations.

\section{PREVIOUS RESULTS}


There are few papers concerning the lowest member of the monophosphate tungsten bronzes family, $m$=2, likely because no phase transition was observed by electrical resistivity measurements and so it was considered unimportant~\cite{Teweldemedhin1991}. However, there is a quasi-one-dimensional character along the direction of the W-octahedra zig-zag chain. The resistivity is about one to two orders of magnitude lower than along the other directions~\cite{Teweldemedhin1991}. A paper of Canadell and Whangbo presents a tight-binding model that predicts a possible electronic instability along the direction of the chain, which is due to the localised and delocalised electrons~\cite{Canadell1990}. 

The structure refinement is reported by Kinomura \textit{et al.}, where two forms of PWO$_5$ are presented. The first form, obtained $via$ 
high pressure synthesis (using 6 GPa), exhibits a tetragonal unit cell with $a$ = 6.25 \AA \ and $c$= 4.07 \AA \ , making it
isostructural to MoPO$_5$. The second form is prepared following the classical protocol used for MPTB. It has 
a pseudo orthorhombic cell with $a$ = 11.172(8) \AA, $b$ = 5.217(1) \AA, $c$ = 6.543(2) \AA, $\upalpha$ = 90$^{\circ}$, $\upbeta$ = 90.34(4)$^{\circ}$ and $\upgamma$ = 90$^{\circ}$ \cite{Kinomura1988}. The structure analysis of this latter form is reported 
by Wang \textit{et al.} as the second member of the MPTBp family  \cite{Wang1989b} and it is the member presented in this paper. Despite the $\upbeta$ being slightly off 90$^{\circ}$, the authors claimed an orthorhombic symmetry~\footnote{This assumption is supported by the low value of the internal reliability factor observed for the $mmm$ Laue class (R$_{int}$ = 0.032)}. In the paper of Wang \textit{et al.}, the non-centrosymmetric space group $Pna2_1$ is considered for the structure solution \cite{Wang1989b}, on the basis of the systematic absence of reflections and statistical tests. The authors reported ``only W atom was refined with anisotropic temperature factors since most of the light atoms gave non-positive definite values''. 
They also signaled high ADP values for three out of five oxygen atoms. The final agreement factor for this 
study is R = 0.0273 for 547 reflections with $I \geq 3 \ \sigma (I)$. However, the slight deviation of the $\upbeta$ angle from 90$^{\circ}$ as well as the difficulty to refine anisotropic ADPs require some clarification, especially since an accurate analysis of the structural state observed below the CDW phase transition has to be performed.

\section{SYNTHESIS}

The method used for synthesis and crystal growth is described by Roussel \textit{et al.}~\cite{Roussel1996}. Normally, a a mixture of (NH$_4$)$_2$HPO$_4$, WO$_3$ and W-precursors in a stoichiometric ratio, leading to the formation of tungsten diphosphate materials. However, an alternative source of phosphorus, P$_2$O$_5$, is required to successfully synthesize the member $m$=2. 
This final mixture must be prepared in a glove box and later put in an evacuated, sealed quartz tube. The ampulla is introduced into a furnace with a temperature gradient.

Crystals of MPTBp $m$=2 with truncated parallelepipedic shape up to millimetric size were isolated. The largest ones were collected for 
physical properties measurements and the small crystals were used for X-Ray diffraction, diffuse and inelastic scattering experiments.

\section{EXPERIMENTAL METHODS}

The quality of the crystals were checked by diffraction in the laboratory and with synchrotron radiation at the European Synchrotron Radiation Facility (ESRF)\footnote{Picture of the sample used for DS and IXS can be found in the Appendix, Fig.\ref{Samplem2}.}. Diffraction data collection was performed using the Synergy-S Rigaku with the Mo microfocus source. A series of measurements for the thermo-diffraction experiment were performed from 290 K to 80K with the same strategy (same sample-detector distances, same current characteristics for the Mo X-ray sources, same exposure time and same value of $\frac{\uplambda}{2 \times \sin (\theta_{max})}=0.8$ \AA \ (\textit{i.e}, $\theta_{max} = 26.315^{\circ}$ \ or $\frac{\sin (\theta_{max})}{\uplambda}$=0.624). The temperature ramp has been set at 1 degree per minute and 10 minutes of stabilization is imposed before each measurement. For each data collection, CrysAlis Pro~\cite{crysalis} was used both for reconstructing oriented diffraction planes from experimental frames and integrating the data. 

The diffuse (DS) and inelastic X-ray scattering (IXS) experiments were conducted on the ID28 beamline at the ESRF. A platelet-like sample of $0.15 \times 0.8 \times 0.2$ mm$^3$ size was used to make temperature-dependence measurements, which were performed with a cryostream (limited to $\sim$90 K). The sample was oriented with the $a$ axis as the rotation axis perpendicular to the beam for the IXS measurement.  A monochromatic beam of $\uplambda=0.6968$ \AA, corresponding to an energy of 17.79 keV, has been used for the DS measurements, whereas the high-energy resolution set-up was required for IXS. The latter is obtained with a beam of 23.72 keV to reach an energy resolution of 1.5 meV.

For the DS measurements, the sample was rotated through 360$^\circ$ orthogonal to the incoming beam and single frames were collected with an angular slicing of 0.1$^\circ$. 
A Pilatus 1M (Dectris) detector with a pixel size of $172 \times 172$ $\upmu$m$^2$ was used in single photon counting mode~\cite{Pilatus}.  
The CrysAlis Pro software package was used to obtain the orientation matrix and to perform a preliminary data evaluation for the following IXS measurements~\cite{crysalis}. The diffuse scattering maps were made through a software locally developed in ID28 called \textit{Project X}.
Further details about the experimental setup of DS and IXS at ID28 have been reported elsewhere~\cite{Girard2019,Krisch2007}.

For resistivity measurements, 4 gold pads were evaporated on a single crystal of size $0.80 \times 0.45 \times 0.40$ mm$^3$. Gold wires were then attached using silver epoxy
Dupont 6828. All measurements were performed in a PPMS (Quantum Design),
using either the resistivity option, external low noise current source
(ADRET A103) and voltmeter (Keithley 2182A) for V(I) curves, and home-made
amplifier/dynamic acquisition card set-up for noise measurements \cite{jojo}.

\section{MOLECULAR DYNAMICS SIMULATIONS}

The molecular dynamics (MD) simulations were done through the $ab-initio$ method. The optimised setup has been reached with a \textbf{k}-point grid of $5 \times 1 \times 3$ and the PBE+GGA (Perdew-Burke-Ernzerhop and generalized gradient approximation) functional has been used to run the simulations. Spin-polarizations have also been considered.
The NVT ensemble was used with a velocity rescaling every 10th step, where the volume is constant and the temperature can be chosen. The thermalization was reached with 1500 steps of 3 fs timestep. The following calculation was run with 9000 steps with a timestep of 5 fs.

\section{RESULTS}

We present our results as follows. In Section 6.1, we discuss the structural results obtained from diffraction measurements performed on both the high-symmetry (HS) phase and the CDW phase. In Section 6.2, MD simulations and symmetry analysis provide insight into the instabilities present in the HS phase, which are linked to the low-temperature CDW phase. Section 6.3 focuses on the pre-transitional phonon instability, investigated using DS and IXS. Finally, in Section 6.4, we examine the electronic transport properties through resistivity measurements, revealing their connection to the CDW phase transition.

\subsection{ Structural analysis}

\subsubsection{Crystal structure in the fundamental state}

A first series of crystals was selected from our own batch synthesis; the chosen samples have parallelepipedic shape and a size of about $100 \times 100 \times 100 \mu m^3$. Data collections were performed using the Synergy-S Rigaku with the Mo microfocus source. The analysis of the reciprocal space leads to the unit cell a = 5.2368(2) \AA, b = 6.5683(2) \AA,  c = 11.2108(3) \AA, $\alpha = 90.217(2)^{\circ}$,  $\beta = 90.000(2)^{\circ}$ \ and $\alpha = 90.001(2)^{\circ}$; the setting chosen here is the one used for all the MPTBp family. As previously reported, a significant deviation of the $\alpha$ \ angle ($\beta$ angle in the setting of by Wang \textit{et al.}~\cite{Wang1989b} from $90^{\circ}$ is evidenced. Moreover, additional weak reflections with $h = \frac{1}{2}$ and $l = \frac{1}{2}$ \ are observed. The introduction of three twin domains related by a three-fold axis parallel to the \textbf{b} axis allows their indexation. 

To eliminate any suspicion related to the quality of the crystals and in particular the influence of twin on both the metric and the refinements, a new crystal of smallest dimensions ($14 \times 41 \times 62 \mu m^3$) but showing sharp Bragg reflections and at first glance the absence of twin was selected. Data collection performed at RT also using the Synergy-S Rigaku with the Mo microfocus source and Eiger 1M Dectris photon counting detector lead to the following unit cell: 
a = 5.22786(5)\AA, b = 6.55814(11)\AA,  c = 11.1883(2)\AA, $\alpha = 90.0583(15)^{\circ}$, $\beta = 90.0118(12)^{\circ}$,  $\gamma = 90.0024(10)^{\circ}$. The deviation of the $\alpha$ angle from $90^{\circ}$ is much smaller for this sample.
It should be noted that low-intensity reflections non-indexed within the chosen unit cell are observed. These features may be interpreted as originating from small crystallites attached to the main crystal, exhibiting a slight misorientation of 2–3 degrees around the a-axis. Representing only about 3\% of the collected reflections, these minor contributions could account for the small deviation of the $\alpha$ angle from the ideal value of 90$^{\circ}$. Although it remains difficult to completely rule out a possible lowering of symmetry toward a monoclinic space group, the overall observations, together with the data integration done using the $mmm$ Laue class leading to a Rint = 0.05 are in favor with an orthorhombic symmetry. The analysis of the diffraction pattern shows the following conditions limiting the diffraction 0k0 : k = 2n ; 00l : l = 2n; h0l : l =2n ; h00 : h = 2n ; hk0 : h+k = 2n leading to two possible space groups : $Pmcn$ or $P2_1cn$. 

\begin{figure}
\caption{
Structure of P$_4$W$_4$O$_{20}$ at room temperature: a) projection along \textbf{a} -the elementary
motif is enlightened by a double pink circle. The first neighbours chains are drawn beyond this pink circle
using the same purple colour as the elementary chain while the second one is light purple tinted; b) View of the (WO6)$_{\infty}$  chains running along \textbf{a}. c) The description of the yaw, on the left, and roll axes, on the right. These axes have libration angles, $\alpha$  and $\beta$, which describe the rigid-body
motions of the octahedra. These oscillations will be described later in \ref{MD}.}
\noindent \begin{centering}
\centering
\includegraphics[width=\textwidth]{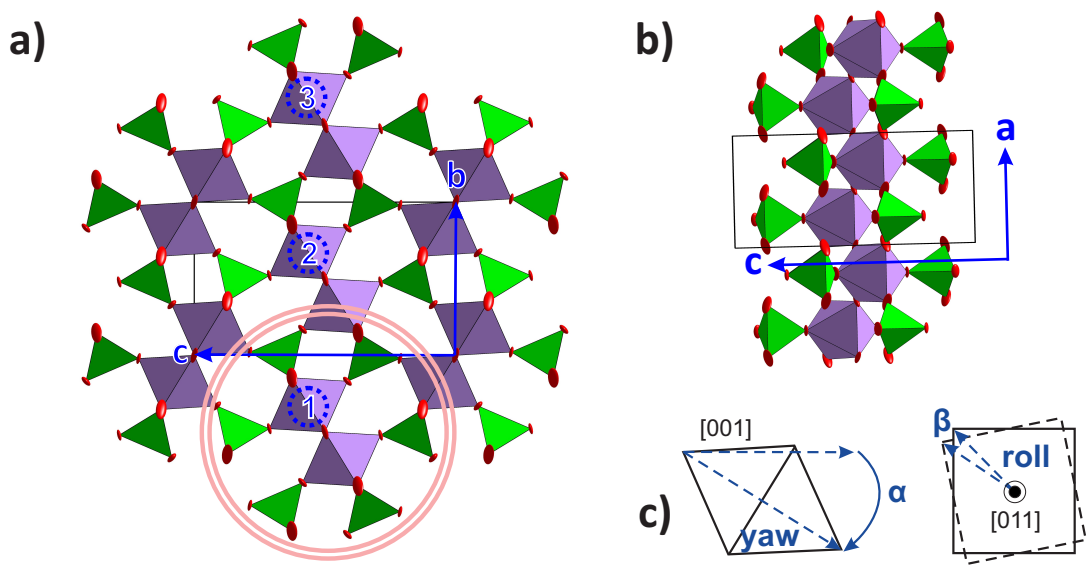}
\par\end{centering}
\label{fig1}
\end{figure}

Reflections statistic (Wilson Plot leads to $<\left|E^2-1\right|> = 0.942$) is in favor of the centro symmetric space group. However, since the results of the literature were in favor of the non centrosymmetric space group, two 	concurrent ab initio structure solutions was performed with the $Pmcn$ and $P2_1cn$. In both cases, following a data integration using Crysalis$^{pro}$~\cite{crysalis}, the structure was solved using superflip~\cite{Oszlanyi2004} and the charge flipping algorithm~\cite{Palatinus2007} and the refinement was performed with Jana2006~\cite{Petricek2014}. 

The structural analysis of $P_4W_4O_{20}$ with the centrosymmetric space group is requiring the introduction of 1 tungsten,
1 phosphorus, and 4 oxygen atoms. The final agreement factor is $R_{obs}= 2.08 \%$ for 952 independent reflections with $I\geq 3 \sigma (I)$  and 41 refined parameters; the ADP of all atoms considered as anisotropic are positive definite. 

In the non centrosymmetric case, the structure is fully described using 1 tungsten, 1 phosphorus, and 5 oxygen atoms. The final agreement factor is $R_{obs}= 2.02 \%$ for 1700 independent reflections with $I\geq 3 \sigma (I)$  and 65 refined parameters but the anisotropic ADP of one oxygen atoms is not positive definite. Moreover two twin domains related by an inversion center have been introduced and the refined twin volumes are equal to 0.50(4)/0.50(4) .  Additionally the ADP reported especially for oxygen atoms  show elongated ellipsoid along \textbf{a} and \textbf{b} but in both cases. The absence of inversion center does not eliminate this anomaly. The atomic parameters corresponding to the two options are proposed as supplementary materials (\ref{Append1} and \ref{Append2}). All these different features are in agreement with the centro symmetric hypothesis.

The structure will be then described using the results obtained with the centro symmetric space group. The figure~\ref{fig1} shows the building principle of the m=2 MPTBp; a pink double circle reveals in the figure~\ref{fig1}a the elementary motif. In the view of this elementary brick projected along \textbf{b} (figure~\ref{fig1}b) a zigzag chain of edge sharing $WO_6$ octahedra (with 6 W-O distances range from 1.84 to 1.97 \AA) running along \textbf{a} can be identified. Each octahedron is connected to 4 $PO_4$ tetrahedra ((with 4 P-O distances range from 1.497 to 1.516 \AA). Consequently, the $(WO_6)_{\infty}$ chains observed in the figure~\ref{fig1}a  and b are isolated ones to each others revealing a 1-D character of the structure of $P_4W_4O_{20}$. Let us signal the analysis of the ADP shows elongated ellipsoids along \textbf{a} and \textbf{b} for the oxygen atoms.

\subsubsection{Thermo-diffraction screening} \label{thermo-dif}

\begin{figure}
	\centering
		\includegraphics[width=\textwidth]{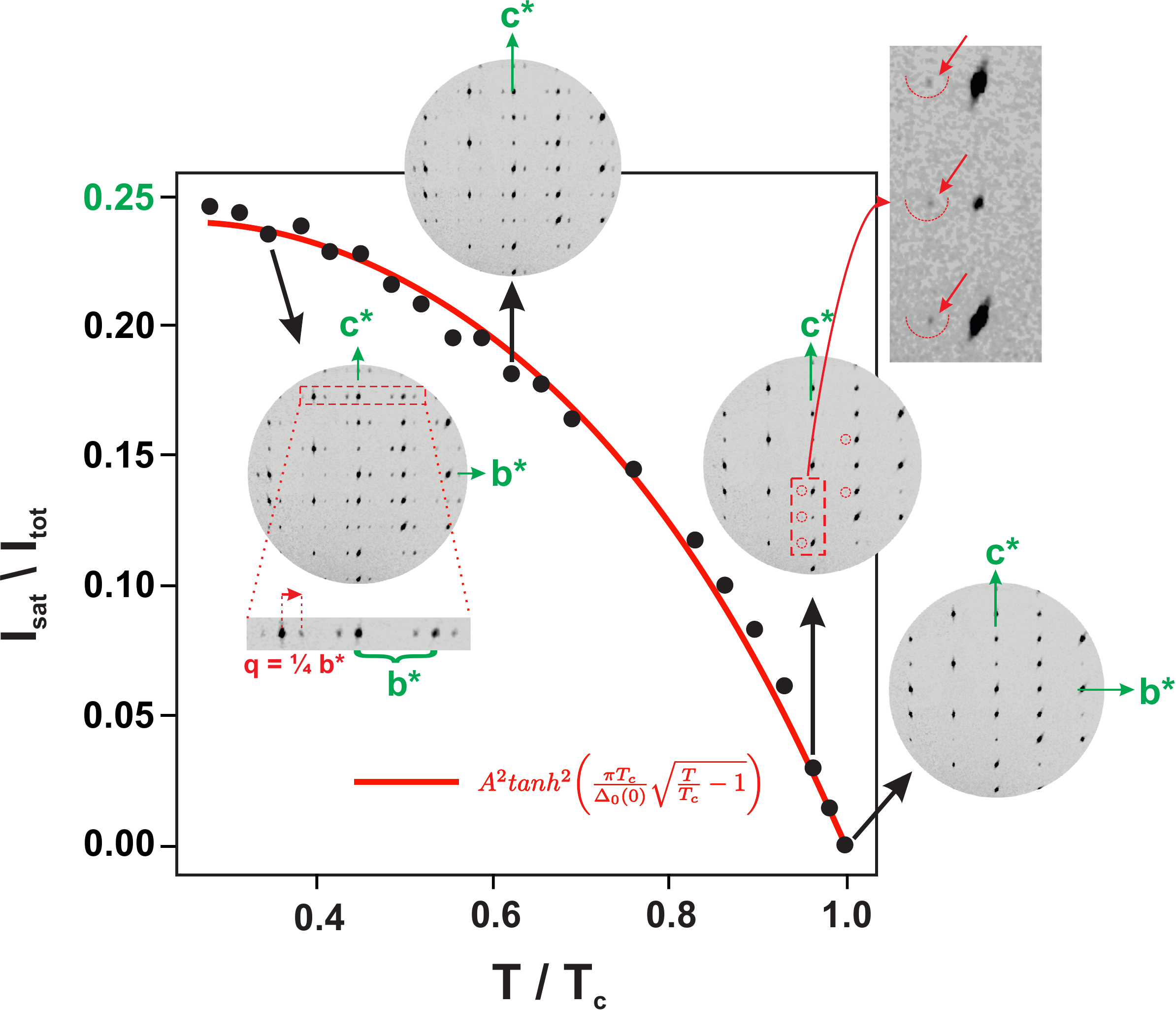}
	\caption{Evolution of satellite intensities versus the total diffracted intensities, as function of the normalised temperature (Tc=290 K). The same area of the 0KL plane is plotted for different temperatures. Below Tc weak satellite reflections are observed (red arrows). The solid line is a fit given by the (squared) expression of the BCS gap in the weak coupling limit ($\Delta_0$/KTc=1.76), \textit{vide infra}.}
	\label{fig2}
\end{figure}

The crystals characterized by the absence of twin and used for the structural analysis at room temperature has been considered for thermo diffraction
experiments. A series of measurement were performed from 290 K to 80K with the same strategy (same sample - detector distances, sames current characteristics for the Mo X-ray sources, same exposure time and same value of $\frac{\lambda}{2 \times \sin (\theta_{max})}=0.8$ \AA \ (\textit{i.e} $\theta_{max} = 26.315^{\circ}$ \ or $\frac{\sin (\theta_{max})}{\lambda}$=0.624). The temperature ramp has been set at 1 degree per minute and 10 minutes of stabilization are imposed before each measurement. 21 data collections were done. For each of them Crysalis$^{pro}$~\cite{crysalis} was used both for reconstructing from the experimental frames oriented diffraction planes and for integrating the data. A selection of regions of (0kl)$^{\star}$ plane are proposed in the figure~\ref{fig2}. The appearance of additional reflections located at $\frac{1}{4}\mathbf{b}^{\star}$ is observed from 280K; the position of these satellites reflections does not evolve with the temperature. The intensity of these satellites reflections is very weak close to the transition temperature but grows up quickly to a pseudo plateau around 80K and seems to reach $\frac{1}{4}$ of the total diffracted intensity (see figure~\ref{fig2} for an evolution of the intensity of satellites reflections the versus T). 

\subsubsection{Crystal structure in the CDW state: Analysis at 80K}

Following the thermo diffraction investigation and to obtain the maximal intensity of the satellite reflections, a full data collection was performed at 80K up to $\frac{\lambda}{2 \times \sin (\theta_{max})}=0.6$ \AA \ (\textit{i.e} $\theta_{max} = 36.234^{\circ}$ \ or $\frac{\sin (\theta_{max})}{\lambda}$=0.832). The data analysis done using Crysalis$^{pro}$~\cite{crysalis} reveals the following unconstrained parameters: 
a = 5.2302(2) \AA, b = 6.5427(5) \AA, c = 11.1823(4) \AA, $\alpha = 89.900(5)^{\circ}$, $\beta = 89.984(3)^{\circ}$, $\gamma = 89.979(5)^{\circ}$  \ and wave vector \textbf{q} =  0.0000(8)\textbf{a}$^{\star}$ +   0.2502(11)\textbf{b}$^{\star}$ +  0.0000(13)\textbf{c}$^{\star}$; the crystal structure can be considered
as commensurate modulated structure. Let us notice the deviation of the $\alpha$ angle is slightly larger than the one identified at room temperature \ldots 
The condition limiting the reflections at 80K h0l: l=2n and hk0: h+k+m=2n in agreement with a c superglide mirror $\perp$ \textbf{b} axis and  $\left( \begin{array}{c}
n \\
s 
\end{array} \right)$ superglide mirror $\perp$ \textbf{c} axis respectively are compatible with the superspace group $Pmcn(0 \sigma_2 0)00s$ and $P2_1cn(0 \sigma_2 0)00s$. The centrosymmetric superspace have been chosen since the space group of the high temperature form is $Pmcn$. 

Data are integrated using Crysalis$^{pro}$~\cite{crysalis}; only main (911 with I $\geq 3 \sigma$ (I)) and first order satellite (1343 with I $\geq 3 \sigma$ (I)) independent reflections are observed. Corresponding the superspace group $Pmcn(0 \sigma_2 0)00s$, the average unit cell contains 6 independent atoms (1 W, 1 P and 4 oxygen). A periodic modulation function expressed as a Fourier expansion developed up to the first order have been introduced to describe the possible atomic displacements of W. The refined atomic displacements are relatively moderate (0.43 \AA \ along y and 0.87 \AA \ along z) but significant; the R factor for the satellite reflections drop to 17.3 \%. One displacive modulation wave are then introduced for P and O atoms; the R factor for satellite reflections reach 6.1 \%. The atomic displacements observed for phosphorus and oxygen are quite large both along y and z (up to 1.06 \AA \ for P and 2.11 \AA \ for O). At this step of the refinement, the different sections of the superspace have to be analyzed; two special sections can be identified:
\begin{itemize}
\item section $t_0=0 + \frac{n}{4}$ \ with the space group $ P 2_1/m \: 1 \: 1$ \ (unique axis \textbf{a}) for the (\textbf{a}, 4\textbf{b}, \textbf{c}) supercell
\item section $t_0=\frac{1}{16} + \frac{n}{4}$ \ with the space group $ P mc2_1 $ \ for the (\textbf{a}, 4\textbf{b}, \textbf{c}) supercell
\end{itemize} 

These two sections are tested, \textit{ie} choosing the commensurate option in Jana2006~\cite{Petricek2014} the structure refinement has been performed using the two possible $t_0$ \ origin along the fourth dimension. A new integration using Crysalis$^{pro}$~\cite{crysalis} but with the \textit{2/m 1 1} Laue class is performed to check the monoclinic section. Consequently, depending of the $t_0$ section, to data sets are used:
\begin{itemize}
\item section $t_0=0 + \frac{n}{4}$, monoclinic symmetry with \textbf{a} unique axis: 
\begin{itemize}
\item 1676 independent main reflections with I $\geq 3 \sigma$ (I)
\item 2466 independent 1$^{st}$ order satellite reflections with I $\geq 3 \sigma$ (I)
\end{itemize}
\item section $t_0=\frac{1}{16}$, orthorhombic symmetry : 
\begin{itemize}
\item 1708 independent main reflections with I $\geq 3 \sigma$ (I)
\item 2427 independent 1$^{st}$ order satellite reflections with I $\geq 3 \sigma$ (I)
\end{itemize}

\end{itemize}   

The refinements were perform in parallel way for the two sections; 67 refined parameters are involved in both cases. 
They converge quickly toward stable solutions and no anomaly in term of both ADP or chemistry (inter-atomic distances 
or angles) is observed. The final R factors are $R_0$ (main reflections) = 2.91 \%  and $R_1$ (first order satellites) =
6.72 \% for the monoclinic section and  $R_0$=2.76  \%  and $R_1$=6.72 for the orthorhombic one.
Unfortunately the difference observed for the agreement factors between the two analyzed sections is too weak to conclude in favor of one or the other solutions.

\begin{figure}
	\centering
		\includegraphics[width=0.9\textwidth]{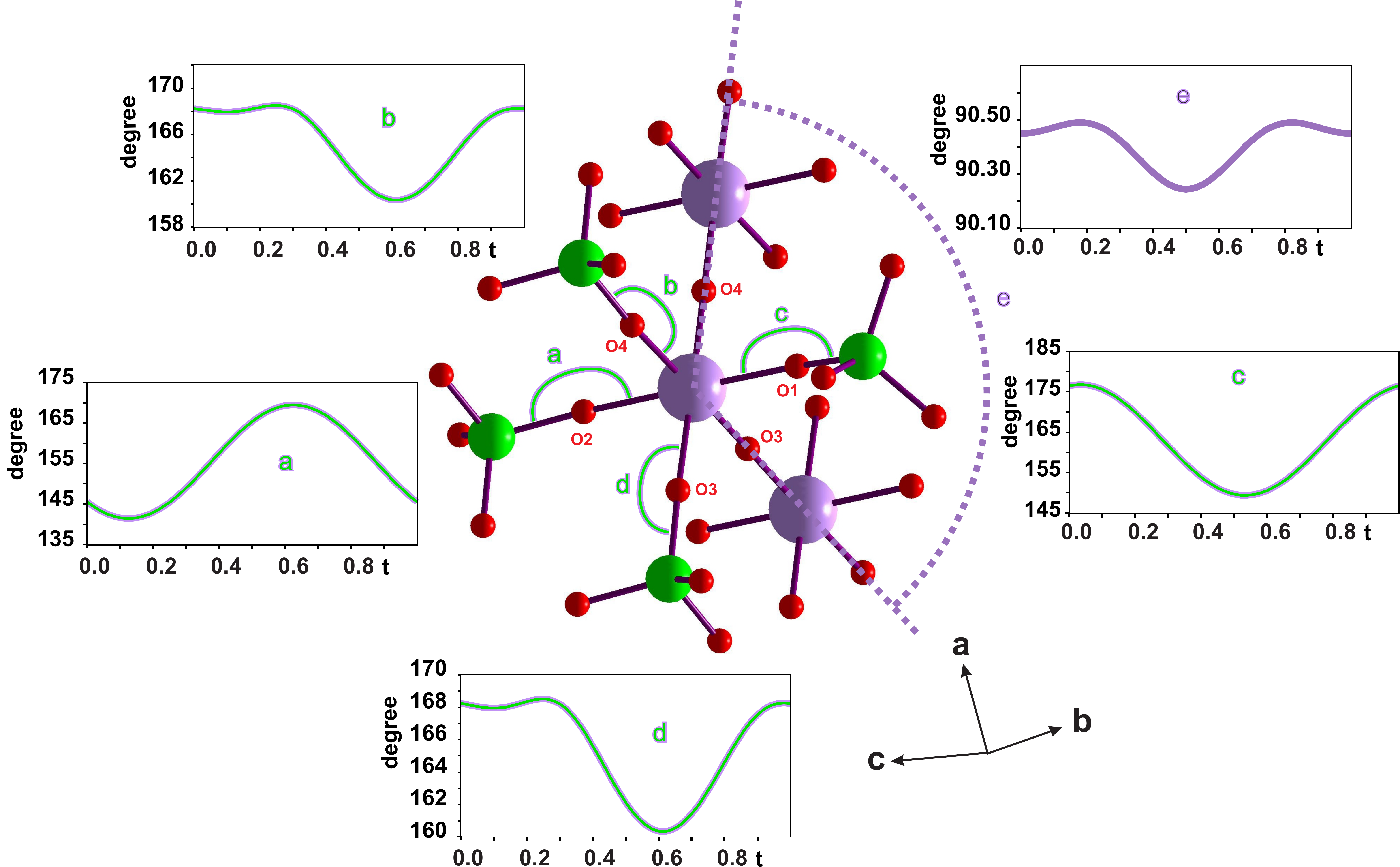}
	\caption{Segment of the (WO$_6$)$_{\infty}$ chains. The evolution, versus t, of the angles noted a, b, c, d, e in the figure are plotted. W, P and O atoms are drawn using purple, green and red color respectively.  }
	\label{figure-5}
\end{figure}

\begin{figure}
	\centering
		\includegraphics[width=\textwidth]{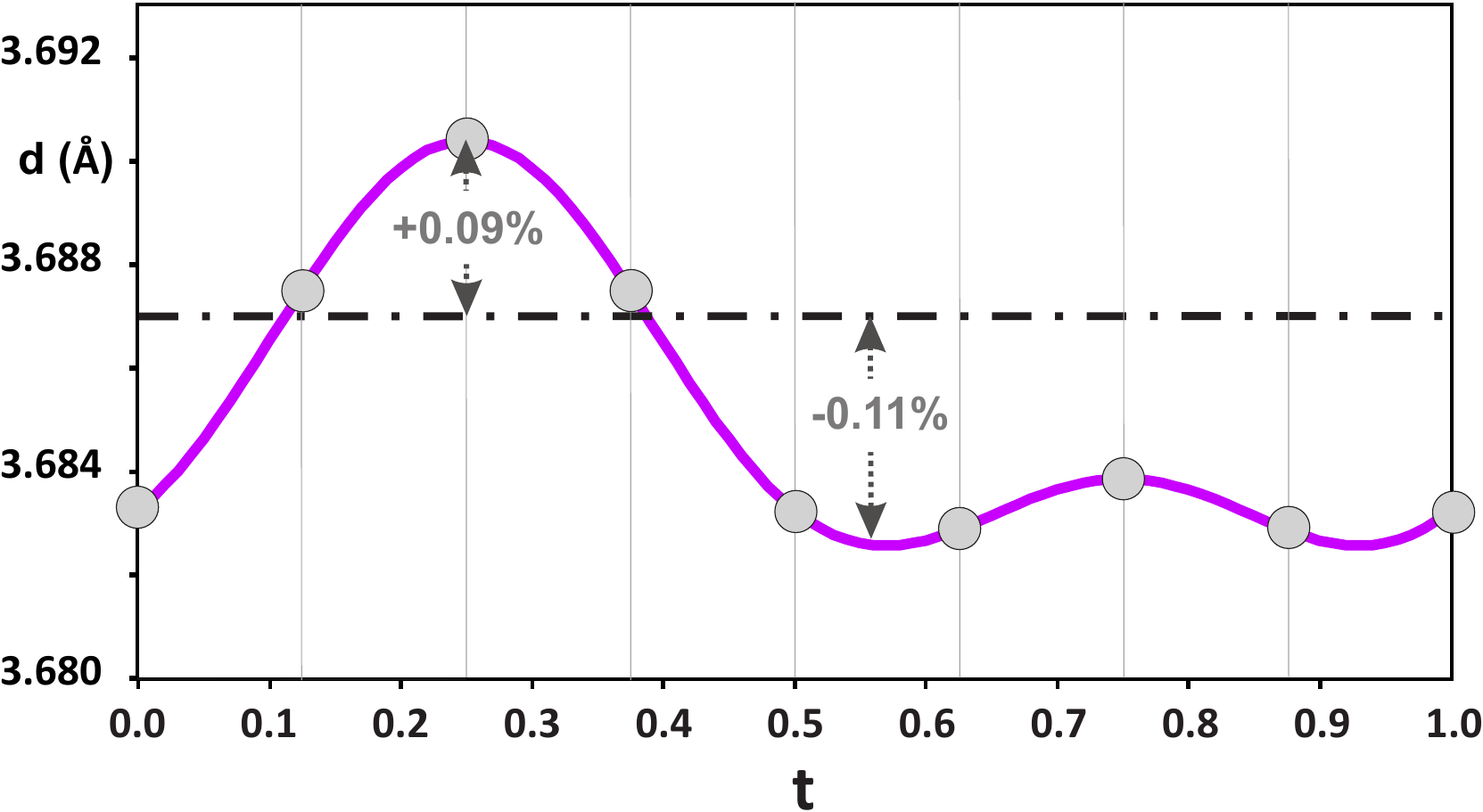}
	\caption{Evolution versus t of the W-W first neighbors within (WO$_6$)$_{\infty}$ chains is proposed. The corresponding distance in the RT structure is drawn using dotted dashed line. Vertical lines
	are showing the relevant distances for the $t_0 = 0$ monoclinic section. }
	\label{figure-6}
\end{figure} 

\subsubsection{High-Q resolution diffraction measurement}

High-Q resolution diffuse scattering measurements at the ESRF were performed over a wide temperature range, spanning from above to below the phase transition (Fig.~\ref{HighQDS}). These measurements reveal several key features that provide valuable insight into the symmetry of both the ground state and the charge-density-wave (CDW) state.

Above the phase transition, the data exhibit pretransitional diffuse scattering signals at positions close to 0.25 \textbf{b*}, while below the transition, well-defined satellite reflections appear at the same positions. The Bragg reflections are saturated over the entire temperature range; therefore, the discussion focuses primarily on the signatures of the phase transition, namely the diffuse scattering and the satellite reflections.

\begin{figure}
	\centering
		\includegraphics[scale=0.8]{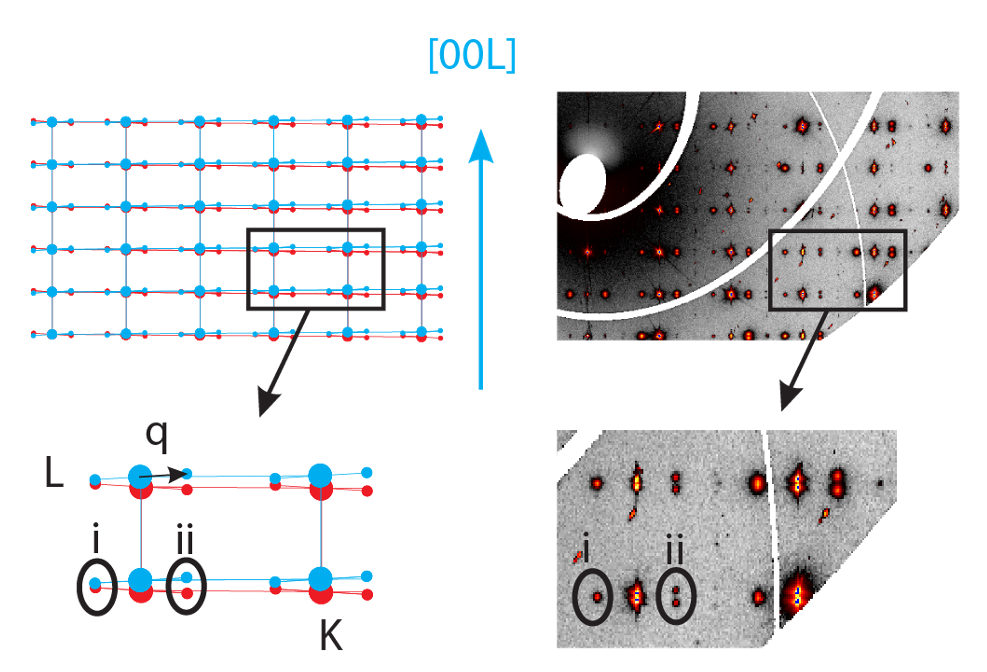}
	\caption{right part: 0KL plane measured at ESRF for DS in the CDW state at 155K. The magnification evidences in the i and ii areas the absence of splitting of the -1 satellite and the splitting of the +1 satellite respectively observed for the high K. left part: model for the 0KL plane built with \textbf{q} = 0.25\textbf{b*} – 0.019\textbf{c*}, the unit cell a = 5.223 \AA, b = 6.548 \AA, c = 11.191 \AA, $\alpha$ = 89.6$^{\circ}$, $\beta$ = $\gamma$ = 90$^{\circ}$, and two domains (blue and red) related by a 180$^{\circ}$ rotation around \textbf{a}. }
	\label{figI}
\end{figure}

Both types of signal show a weak but systematic broadening along the c* direction. This broadening is consistent with the presence of a monoclinic distortion and the existence of twin domains already in the ground state. The effect becomes more pronounced upon decreasing temperature (see Fig.~\ref{HighQDS}). Figure~\ref{figI} presents data collected below the phase transition at 155 K. The right-hand side of Fig.~\ref{figI} shows the 0KL diffraction plane, where both the main Bragg reflections and the satellite reflections are clearly visible. As k increases, the main reflections broaden anisotropically along \textbf{c*}. Even more strikingly, the +1-order satellite reflections exhibit an increasing splitting with increasing k, whereas the ?1-order satellites do not show any comparable splitting.

These observations strongly suggest the presence of a twinned domain structure, most likely originating from a lowering of symmetry toward a monoclinic Laue class. To quantify this monoclinic distortion, we analyzed the broadening of the main reflections and obtained an angle $\alpha$ of approximately 89.6$^{\circ}$. However, neither this angular deviation nor the presence of a twin domain related by a 180$^{\circ}$ rotation around \textbf{b} or \textbf{c} can account for the asymmetries observed in the satellite reflections (e.g., those labeled i and ii in right part of Figure~\ref{figI}).

To explain this feature, it is necessary to introduce an additional component along \textbf{c*} into the modulation vector. The best agreement with the experimental data is obtained when a component of 0.019\textbf{c*} is included. A diffraction pattern schematized with \textbf{q} = 0.25\textbf{b*} + 0.019\textbf{c*}, cell parameters a = 5.223 \AA, b = 6.548 \AA, c = 11.191 \AA, $\alpha$ = 89.6$^{\circ}$, $\beta$ = $\gamma$ = 90$^{\circ}$, and two domains related by a 180$^{\circ}$ rotation around \textbf{a}, reproduces the observed diffraction plane remarkably well (left part of Fig.\ref{figI}), including the characteristic behavior of satellites i and ii.

This two-component modulation vector, incompatible with the $mmm$ Laue class, therefore strongly supports a monoclinic symmetry.

$P_4W_4O_{20}$ thus appears to possess monoclinic symmetry both above and below the phase transition. Our laboratory single-crystal X-ray diffraction data collected from room temperature down to 80 K do not reveal any explicit signature of symmetry lowering, as they do not allow the weak monoclinic distortion or the small \textbf{c*} component of the modulation vector to be resolved. Nevertheless, the refinement presented in the previous section particularly the $t_0$ section of the superspace description leading to the space group $P2_1/m11$ for the supercell is fully consistent with both the deviation of $\alpha$ from 90$^{\circ}$ and the \textbf{b*} and \textbf{c*} components of the modulation vector.

The structural refinements proposed above and below the phase transition therefore provide a reliable approximation of the structure of MPTBp ($m$ = 2) in both the ground state and the CDW state. Moreover, this structural model yields improved agreement with molecular dynamics simulations, and it is this approximation that we describe in the following section.

\subsubsection{Description of the structural model}

\subsubsection{analysis in the $ 80K \leq T \leq 285K $ interval}

\begin{figure}
	\centering
		\caption{Evolution of the maximum atomic displacements extracted from figure~\ref{figure-SI1} for W1 (purple) and O4 (red) versus T.}

		\includegraphics[width=0.8\textwidth]{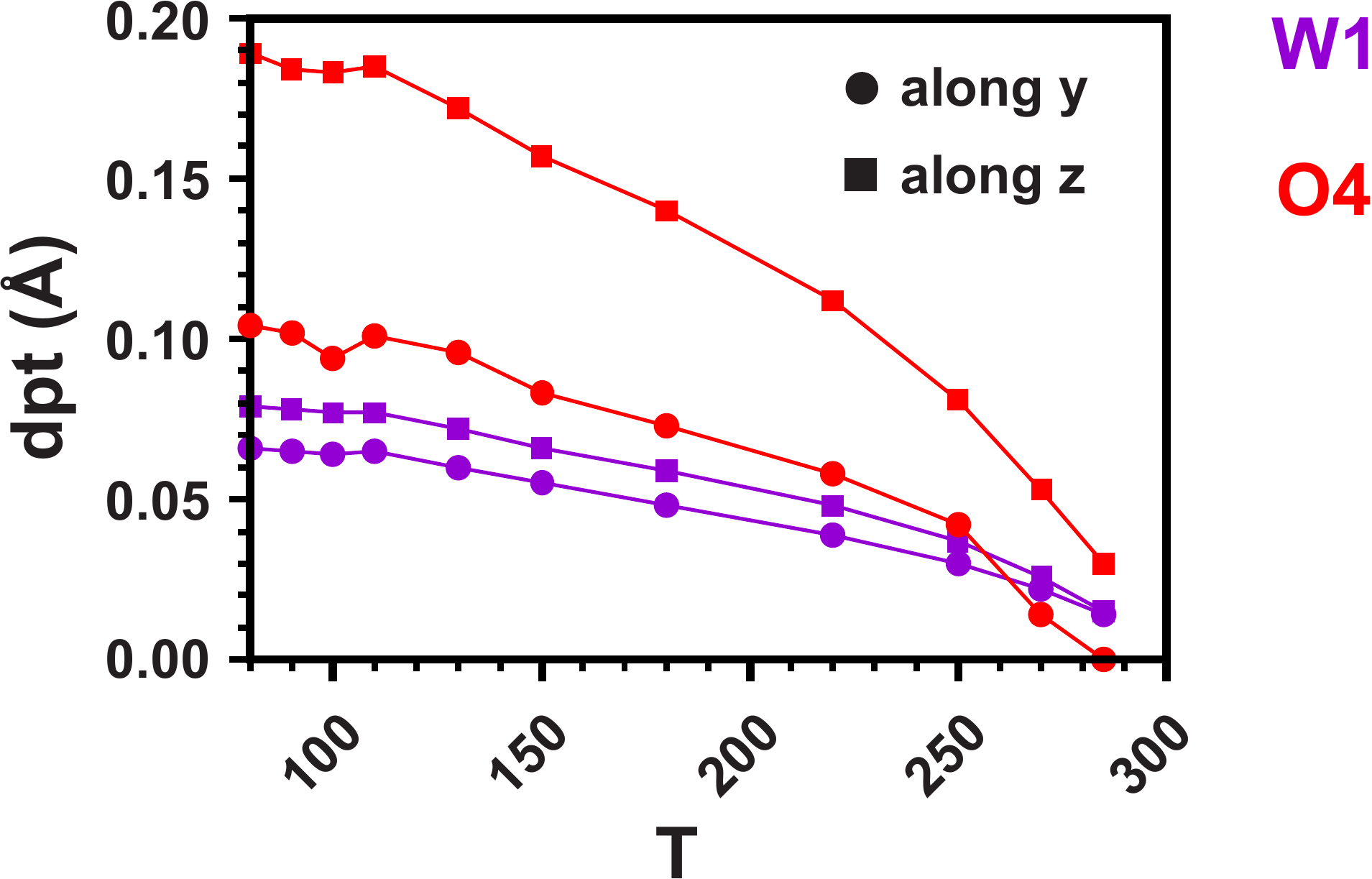}
	\label{figure-7}
\end{figure}

Following the study carried out at 80K, an analysis of a sampling of the data measured between 80K and 285K was carried out. The aim of this analysis was to evidence the impact of this decrease on the modulation and especially the maximum atomic displacements and then to follow the modulation versus T. The data at 80K, 110K and 150K were collected up to up to $\frac{\sin (\theta_{max})}{\lambda}$=0.832; data set at 130K, 180K, 220K 250K, 270K, 285K are limited to $\frac{\sin (\theta_{max})}{\lambda}$=0.624. The different parameters used to validate the structural refinements are satisfying. Nevertheless the reliability factor calculated for the first order satellite reflections at 285K is relatively high ($\approx$ 26 \%). But as previously reported in the  \ref{thermo-dif} part, the intensity of the satellite reflections decreases when the temperature increases and then for this data set only a few satellites with very weak intensity are observed (120 first order satellites corresponding to 1.4\% of the global diffracted intensity at 285K against 881 first order satellites corresponding to 22.8\% of the global diffracted intensity at 130K). Be that as it may, the main characteristics of the modulation can be followed at the different temperatures. The figure~\ref{figure-7} shows the evolution of the atomic displacements for the tungsten and one oxygen atoms; it can be observed that both along y and z the displacement increases as T decreases. The maximum displacements for these two atoms can be extracted at
the different temperatures; they are plotted in the figure~\ref{figure-7}. The shape of the different curves are very similar to the evolution of the satellite intensity versus T (fig.~\ref{fig2}).

\subsection{ MD simulations $\&$ symmetry analysis}\label{MD}


The molecular dynamics (MD) show interesting results, especially for the high temperature phase. We introduce the concept of axis librations, which will be useful for understanding the following part. The libration is the oscillation of the axes of rotation with an inclination angle linked to the ``normal'' axes. The librations are considered for the 4-fold axes of the tungsten octahedra, as they were found the most prominent movements. The three rotational axes of the octahedron are called yaw, roll and pitch and the first two are described in figure~\ref{fig1}c. The yaw axis runs parallel to the [100] direction, whereas the pitch and the roll axes are in the YZ plane. The pitch axis is parallel to the [110] axis and the roll axis is perpendicular to it. Those oscillations describe a preponderant displacement of the oxygen atoms of the octahedra.\\

\begin{figure}
\includegraphics[width=\textwidth]{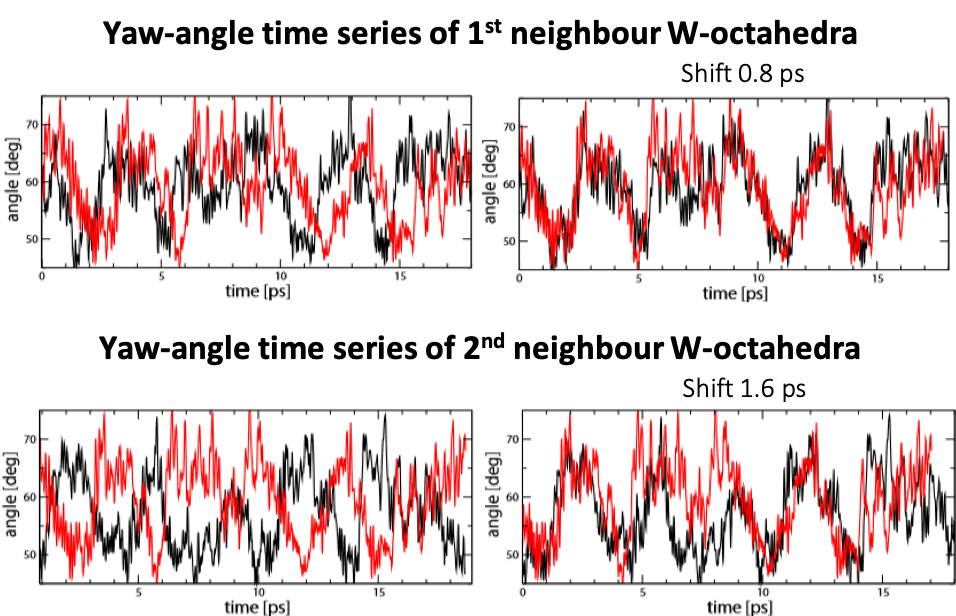}
\caption{Time dependence of the yaw angle values for the two octahedra of different chains along $b$, as a result of the MD simulations at 300K. On the top, the angle time series before and after a time delay of the 1$^{st}$ chains-neighbours octahedra (black and red line for each one), on the bottom, of the 2$^{nd}$ chains-neighbours octahedra (black and red line for each one). A shift is applied to show the high correlation between these time series at 0.8 ps and 1.6 ps, respectively. The 1$^{st}$ and 2$^{nd}$ neighbours octahedra are marked with the numbers 1,2 and 3 in figure~\ref{fig1}a, whereas the visualization of the yaw axes libration and its large angle of libration can be seen in figure~\ref{fig1}c. }
\label{yawaxis}
\end{figure}

The superstructure modulation of the CDW phase can be described by the yaw axis libration, which is the predominant oscillation. The librations last 3-4 ps with a magnitude of $\sim 12 ^{\circ} $ of the average angle, which is about $ 62 ^{\circ} $ with respect to the [001] direction. Compared to the other two modes, it is larger and slower~\footnote{In fact, the roll axis libration is a few times faster with a short-range correlation and its angle is smaller. The libration of the pitch axis can be seen as thermal noise, since it is high frequency.}. In fact, the octahedron rotation around the [100] axis has a change of phase by 2$\pi$/4 with respect to the neighbour W-octahedron in the [010] direction. The rotations in time and space are shown in Fig. \ref{yawaxis} for the simulation at room temperature. The scheme contains the yaw-angle-libration time series of the W-octahedra related by (010) and (020) translations. With the octahedra connected by a (010) translation, the high correlation is shown thanks to a 0.8 ps shifting, whereas for the two connected by a (020) translation, the shifting is doubled. It demonstrates that the yaw axis vibrations break the (010) translation period. This result perfectly explains the 4-fold supercell, and we can also conclude that the principal degree of freedom is based on the rigid-body rotations of octahedra, which corroborate the diffraction results.  \\

\begin{figure}
\caption{The two principal modes of the ``order-disorder'' structural phase transition, LD4 and GM1. The LD4 mode can be seen as an anti- and clockwise movement of the octahedra, drawn with a black and white circular arrow, respectively. The GM1 mode is an opposite pair movement of the oxygen atoms (straight red arrow). The modes were found by ISODISTORT~\cite{Campbell2006, Isodistort} looking at the structure refinements for the high-temperature phase and the CDW ground-state, a monoclinic structure. }
\includegraphics[width=0.5\textwidth]{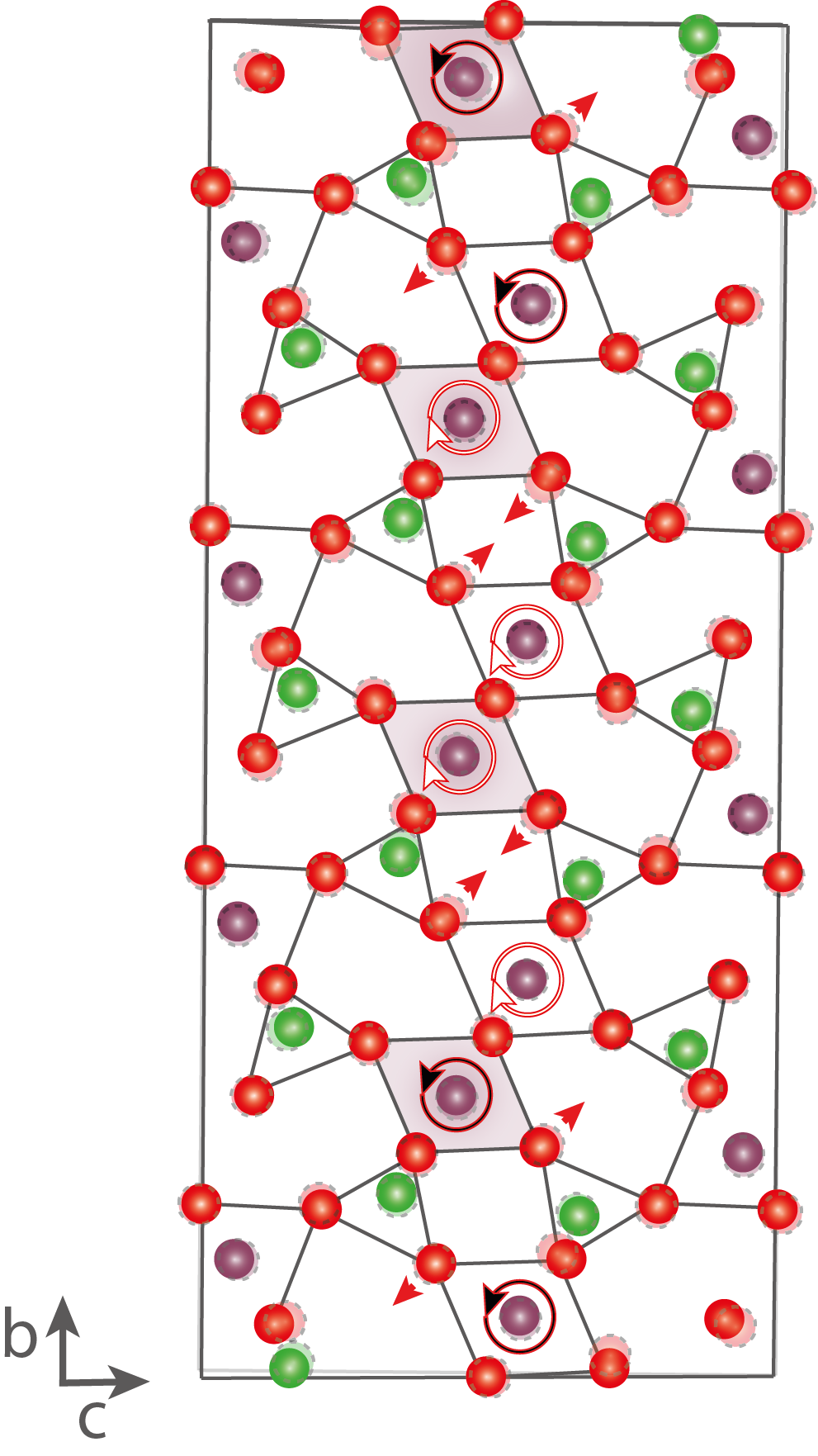}

\label{Mode}
\end{figure}


With those results to add to the diffraction results shown before, the global picture seems to consider the structural phase transition as ``order-disorder''. The structure at high temperature is almost orthorhombic and, as reported earlier from the diffraction results, it transforms at the CDW ground state to a monoclinic phase. Two modes describe the transition, LD4 and GM1. The first one is the primary mode with a k-vector along \textbf{b*}, and the latter is coupled with the primary. The two modes are described in figure \ref{Mode}.  Choosing the monoclinic CDW ground-state option, the two modes create the movement as described from the MD simulations, the so-called axes librations (which is exactly the sum of the two modes found by the rigid-model). This movement can be seen as a strong atomic displacement of the oxygen and phosphate atoms. It describes the rigid body movement of the octahedra driven by oxygen atoms freedom. The rigid body motion gives rise to a little readjustment of the tungsten atom position. \footnote{The modes and the collective movement given by the phase transition can be seen with the Isoviz file created by ISODISTORT. The file can be found in attachment to the paper.}

\subsection{ Dynamics-related measurements (DS \& IXS)}

\begin{figure}
\caption{Diffuse scattering 0KL maps at different temperatures, above the transition temperature (T$_{CDW}$+180K and +90K), RT and at the transition temperature. A focus of (01$\overline{6}$) shows the part of reciprocal space further studied with the inelastic scattering. It also visualizes the diffuse and Bragg peak satellite at \textbf{q} $\sim$ -0.25\textbf{b*}.}
\includegraphics[width=\textwidth]{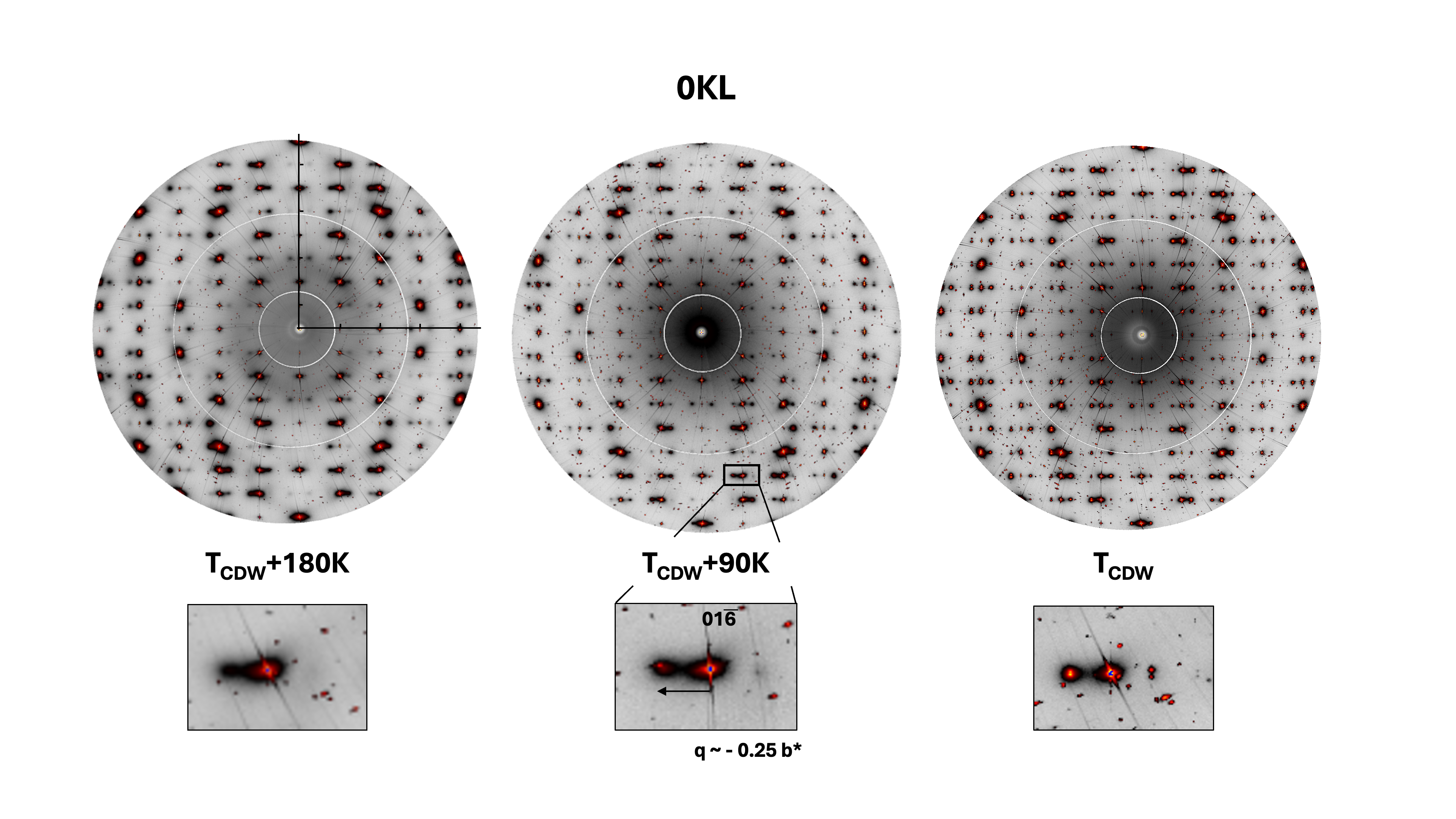}

\label{DS-T}
\end{figure}


\begin{figure}[]
\caption{The projections on the three directions (a,b$\&$c) around 0,0.75,$\overline{5}$ above and below the transition. The experimental data are shown by dots, whereas the fitting of two temperatures are shown by dotted lines, as an example of the high-T phase (T$_C$+200K) and of the CDW ground-state phase (T$_C$-20K) peaks. In $c^*$, the twinning was considered, with the distance and intensity ratio between the two kept constant. The temperature dependence of 1/$\Delta$Q on the three directions show the isotropic shape of the diffuse pattern, reaching full periodicity at the transition temperature.}
\includegraphics[width=\textwidth]{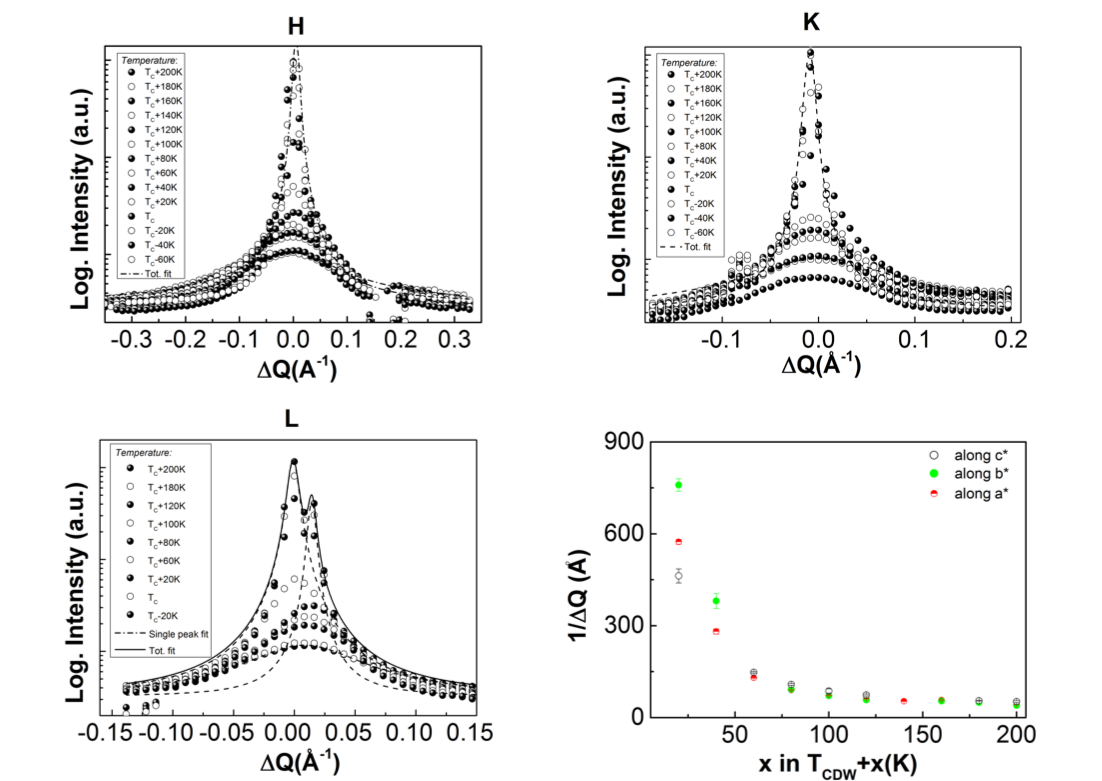}

\label{Proj-DS}
\end{figure}

The inelastic measurements were done on the high-symmetry phase approaching the CDW ground state, in order to reveal the Kohn anomaly and the phonon behaviour linked to the electronic/structural instability. The maps of diffuse scattering on the H0L plane show a temperature dependence, where a pre-transitional diffuse pattern is isotropically distributed around the \textbf{Q}-point related to the satellite of the CDW phase, as shown in figure~\ref{DS-T}. In fact, the distribution is positioned about the centre of the CDW-vector, with a \textbf{b*} component of $\sim 0.25$, considering the Bragg's node. In order to study the temperature dependence of the diffuse, the chosen \textbf{Q}-point for the DS measurement should be completely separated from the Bragg's node. Moreover, a weak Bragg does not present any strong thermal component and the temperature-dependence of the pre-transitional diffuse on the CDW satellite can be better observed. For this reason, we choose the CDW satellite at the position $\sim 0,0.75,\overline{5}$. The shape is isotropic and the evolution in temperature is similar for all of them, as shown in figure~\ref{Proj-DS}. \footnote{The complete temperature evolution of the satellite in the three directions can be followed in the 0KL and the H0.75L DS maps shown in fig.~\ref{Maps-diffuse} in the Appendix}.

The projections of the diffuse in the three directions permit the evaluation of the temperature dependence of 1/$\Delta$Q approaching the transition (fig.\ref{Proj-DS}). This can be extracted with a fit of the peak through a Voigt function, where the instrumental resolution is described through a Gaussian profile and the diffuse through a Lorentzian. The background is fixed with a constant, $y0$, which was taken by averaging a segment of reciprocal space where no Bragg or diffuse are present. From the Lorentzian Full-Width-Half-Maxima (FWHM), we confirm similar behaviour in the three directions, reaching the resolution function around the transition temperature, as shown in table~\ref{CorrLeng}. As expected, at the transition temperature, the diffuse scattering signal is becoming completely elastic following the periodicity of the full crystal. Thus, the Bragg satellite is shaped in each direction.\\ 

\begin{table}
\centering
\begin{tabular}{cccc}
\hline \hline \\
\large {\textit{T} (K)} &	\large {\textit{ $1/ \Delta Q_{a*}$ }} (\AA)  & \large {\textit{$1/ \Delta Q_{c*}$}} (\AA)  & \large {\textit{ $1/ \Delta Q_{b*} $}} (\AA) \\ [0.5ex]
\hline \\
T$_c$+200 & 43.0(6) & 40(2) & 52(3)   \\
T$_c$+120 & 57(2)  & 54(2) & 74(3)  \\
T$_c$+80  & 89.6(1.4) & 92(2) & 108(3)  \\
T$_c$+30 & 574(8) & 759(20)  & 462(23)  \\
\hline \hline \\

\end{tabular} \\
    \caption{Temperature dependence of 1/ $\Delta$ Q along \textbf{a*}, \textbf{b*} and \textbf{c*}.}
    \label{CorrLeng}
\end{table}

Due to the temperature shift caused by sample heating during diffuse scattering measurements (approximately 100~K), additional rocking curve scans were performed using the long-arm spectrometer at beamline ID28. The rocking curves collected at various temperatures are presented in the Appendix, shown in fig.~\ref{Rockingcurve}. From these measurements, we confirm that the elastic component of the CDW phase clearly emerges around 290~K. The data fitting procedure follows the same approach used for the diffuse scattering analysis. In the pre-transitional state, the satellite peak is well-described by a Lorentzian profile. As the system approaches the transition temperature, the instrumental resolution, modeled by a Gaussian profile, must be taken into account. For this reason, a pseudo-Voigt function is used in this intermediate regime. Below the transition temperature, the peak shape is dominated by the instrumental resolution, and a Gaussian profile is sufficient to describe the data. The FWHM of the fitting at different temperatures is reported in the Appendix in fig.~\ref{FWHM} The transition temperature derived from this analysis is approximately 280~K, about 10~K lower than the value obtained from diffraction measurements. This discrepancy may arise from instrumental uncertainties, as both measurements were conducted using a cryogenic gas stream. \\

\begin{figure}[]
\includegraphics[width=\textwidth]{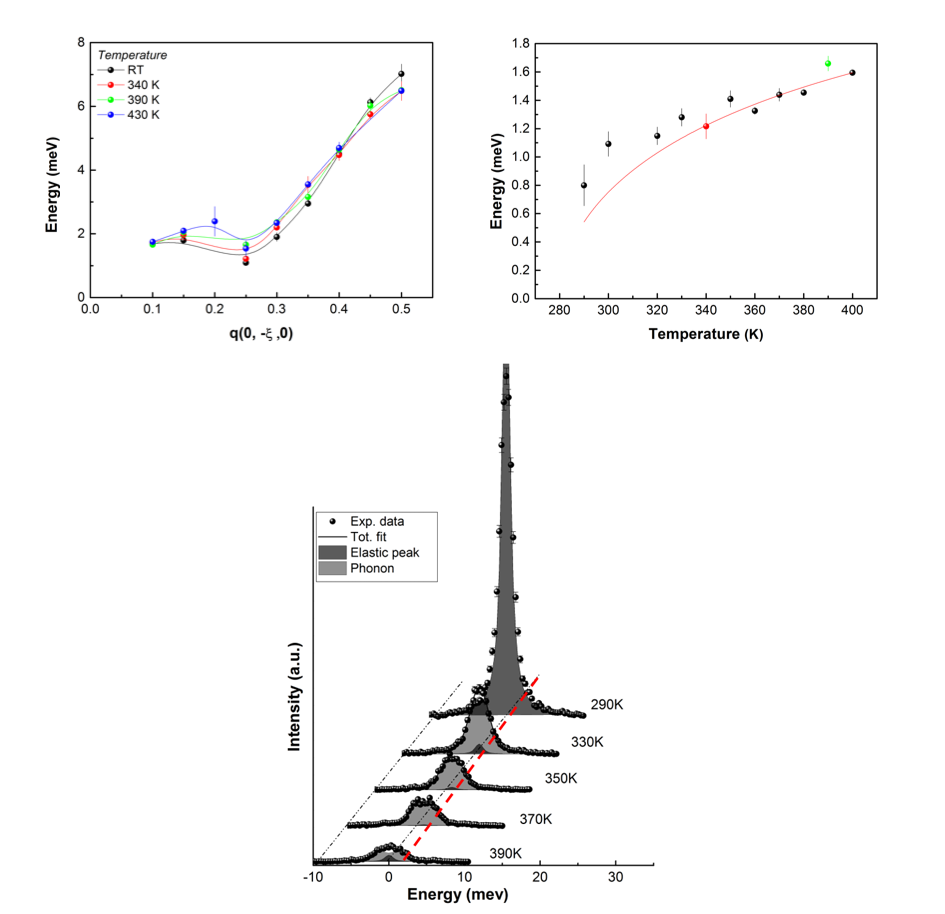}
\caption{In the figure on the top-left the full dispersion from the Bragg peak to the border of the Brillouin zone, at the position Y, in the negative direction $[ 0 -\xi 0 ]$. On the top-right, the temperature dependence of the energy on the satellite position is shown and some of the IXS scans on 0,0.75,$\overline{6}$ are shown in the figure at the bottom, with the fitting. In the latter, we can see the strong elastic component already at 290K.}
\label{Softening}
\end{figure}

Following the position of the CDW and the pre-transitional diffuse pattern, the phonon dispersion along the \textbf{b*} direction from the $01\overline{6}$ Bragg peak was measured through IXS.
In this case, to measure the dynamic component, we chose an intense Bragg. This insures we observe intense phonon branches simultaneously. The ROI (region of interest) in reciprocal space can be seen in the zoomed inset of the DS maps at three different temperatures in figure \ref{DS-T}. 
The satellite is condensing, and a clear phonon softening is present at \textbf{q$_{CDW}$}. The full dispersion has been measured at four different temperatures cooling the system, as shown in figure \ref{Softening}a.  The temperature dependence of the phonon energy is reported in figure \ref{Softening}b, where the single inelastic scans were recorded from 400K to 290K, figure \ref{Softening}c. The energy drops from E=(1.66(50))meV at 390K to E=(0.80(15)))meV at 290K. At lower temperature, the elastic contribution is too strong, hiding the inelastic signal, as it is shown in figure \ref{Softening}c. Expecting the transition at $\sim$280K and using a power law function, the phonon does not seem to soften to zero energy. The fitting function is E=E$_0$(1-T$_{CDW}$/T)$^{\upgamma}$ to the power of 0.5, as suggested for a second order transition within the mean-field theory~\cite{Chaikin1995, Scott1974} , and E$_0$ corresponds to the high-T energy limit. The fitting function is described by the red line in figure \ref{Softening}b. The behaviour does not follow the predicted temperature dependence of a Kohn anomaly with a complete freezing of the vibration as found for other CDW systems~\cite{Hoesch2009,Pouget1985b,Pouget1991}.

\subsection{ Electronic transport properties and relevance of a CDW}

\begin{figure}
\includegraphics[width=\textwidth]{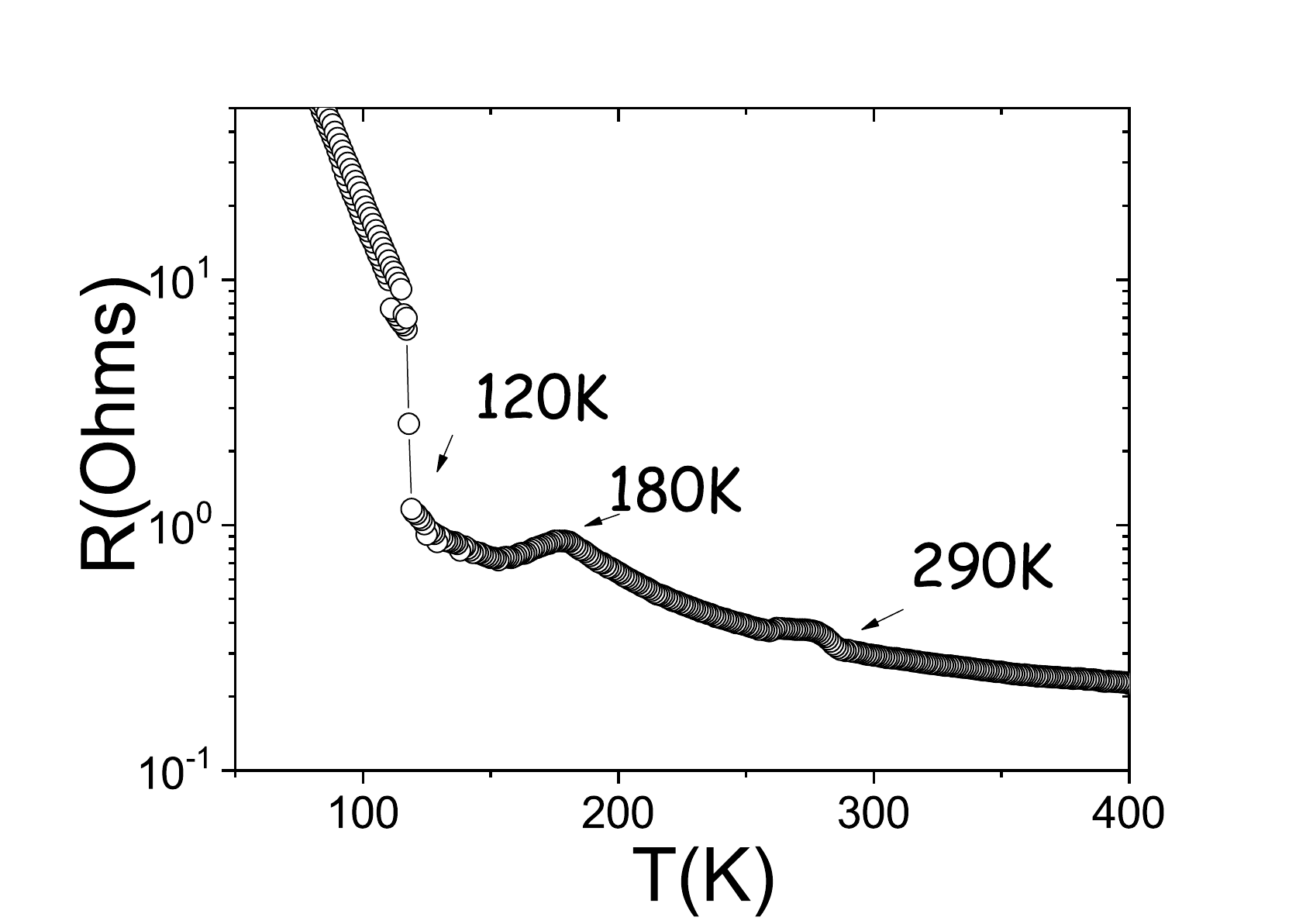} \caption{Resistivity versus temperature measured with a biased current with
I=1mA}
\label{resistivity} 
\end{figure}

Previous measurements have
shown a semiconducting-like resistivity in the 50-400K range \cite{Teweldemedhin1991},
with no evidence of electronic instability. Our new data indicate more complex resistive behavior, as reported in figure \ref{resistivity}.

\begin{figure}
\includegraphics[width=1\textwidth]{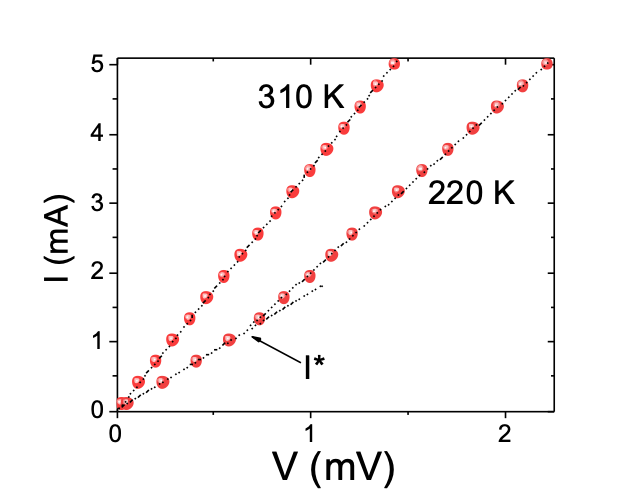} 
\caption{I(V) curve measured at two different temperatures. It should be noted the Ohmic (resp. non Ohmic) behavior at T= 310 (resp. 220) K.}
\label{IdeV} 
\end{figure}

When cooling the sample from room to low temperature, measuring
with a current of 1 mA, three anomalies are depicted
at T$\sim$ 290, 180 and 120K in resistivity. A current dependence study
reveals that the two anomalies at lowest temperature are strongly
shifted by a merely moderate current (typically 100- 1000 $\upmu$A).
It evidences a dominant non-linear behavior. We have then performed
isothermal V-voltage I-current measurements from 320K to 70K with
steps of 5K, using an external set-up (current source ADRET A103, Nanovolmeter
Keithey 2182A). Some typical curves are shown in figure \ref{IdeV}.

\begin{figure}
\includegraphics[width=\textwidth]{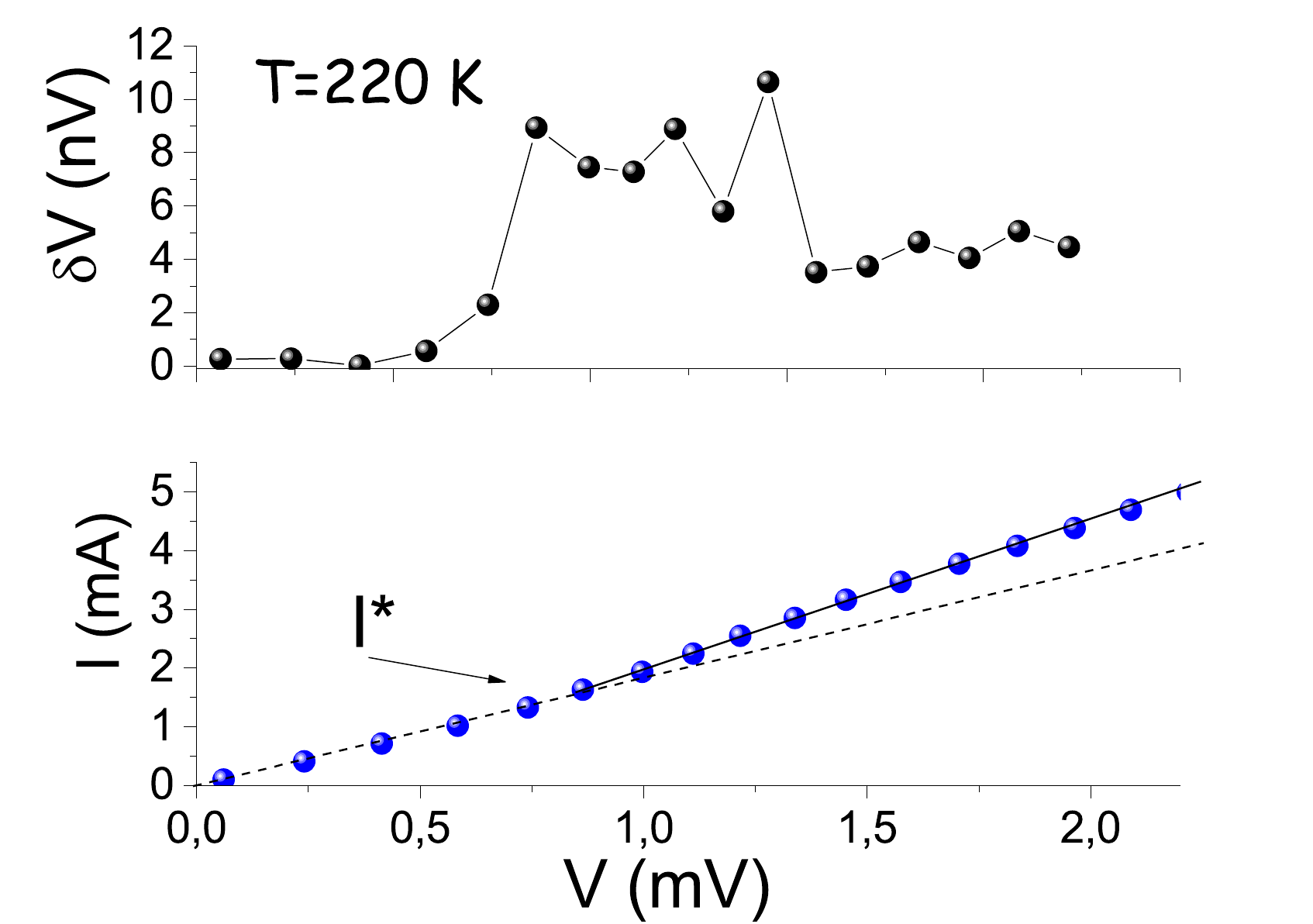} \caption{Top: $\delta V$ as function of applied current, showing a strong increase at $I \sim I^{*}$ which is typical of a depinning transition. Bottom: V(I) curve measured at $T=$220K, with $I^{*}$ the threshold current marking the non-linear behavior.}
\label{noise} 
\end{figure}

\begin{figure}
\includegraphics[width=\textwidth]{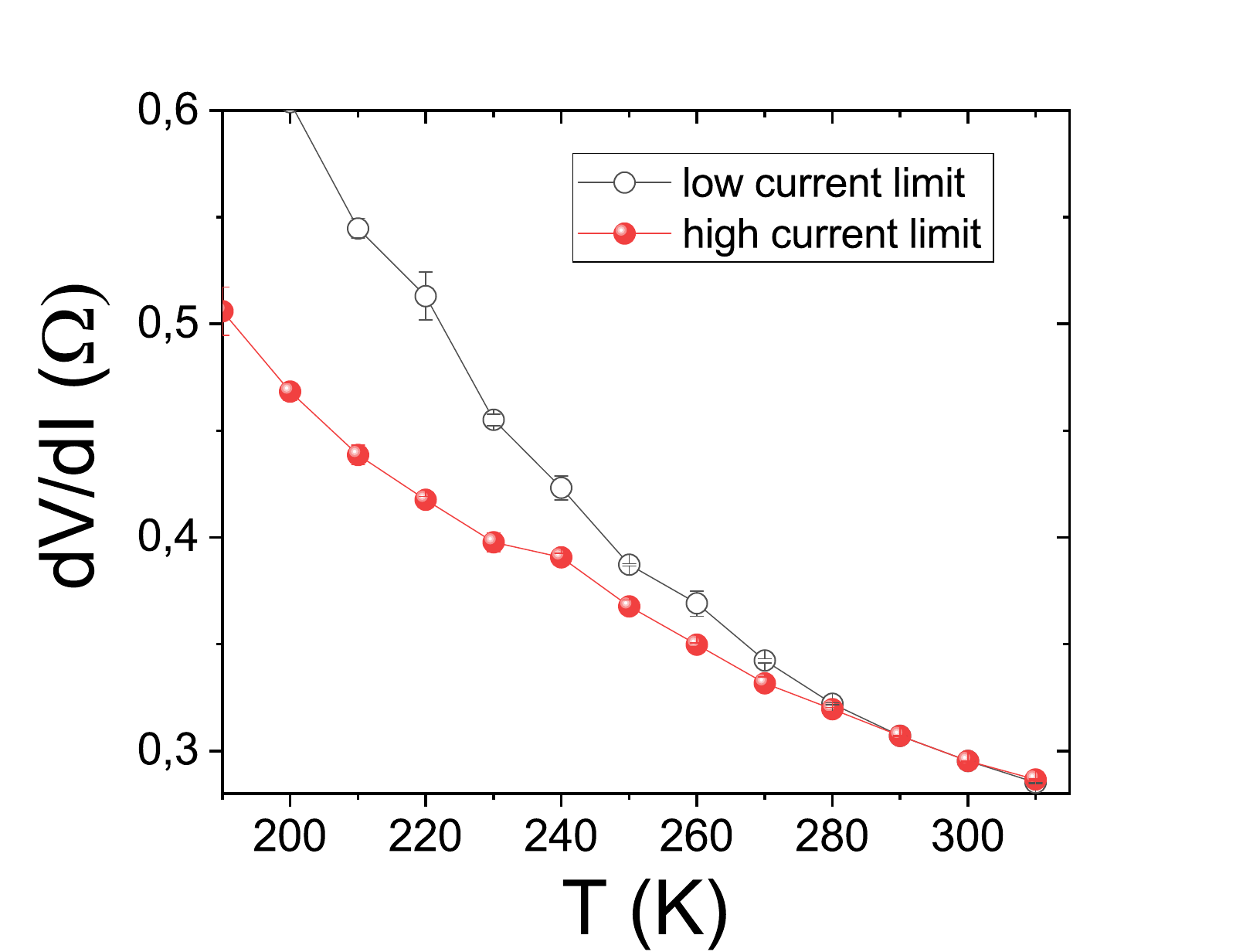} \caption{Differential resistance, deduced from the slopes of V(I) curves at low and high current as function of temperature. Slopes are different, \textit{i.e.}, non-ohmicity appears for T between 280 and 290 K. }
\label{nonlineardeT} 
\end{figure}

In the high temperature range, the V(I) curve is that of an ohmic,
conventional conductor with a single slope R=V/I. At T$\lesssim$
280K, a slight departure from a perfect ohmicity can be noted, which
becomes more clear as the temperature decreases. At T$\lesssim$ 240K,
a two-slopes regime is obvious and evidences an excess of flowing
current above a threshold value noted I$^{*}$. We have fitted the V(I) slope in the high and low current limits to find the temperature where non-ohmic behavior appears. The values at each temperature are reported in figure \ref{nonlineardeT}). Within resolution, the two slopes differ at temperatures below T=280-290 K.
 This behavior is characteristic
of pinning/sliding collective modes of charge density waves \cite{monceau},
as reported for example in the quasi-1D conductor NbSe$_{3}$. In
addition to non-linear conductivity, sliding CDW are expected to generate
both low frequency broadband noise (BBN) \cite{BBN} mainly due to CDW configuration and velocity
fluctuations and high frequency narrow band noise (NBN) \cite{shobo, weissman} due to its
harmonic nature. Here, we have not attempted to measure the NBN, but
focused on the BBN which is expected to be, as a general rule for
depinning transition, large close to the threshold field. At a fixed
temperature, for each value of (noise-free) applied current, we have
recorded the autopower spectra S$_{VV}$(f) ($V^{2}$/Hz) of the voltage
noise, then calculated the noise power defined here by $\delta V^{2}$=${\displaystyle \int_{0.1}^{10}S_{VV}(f)df}$.
As expected, $\delta V^{2}$ does not depend on the applied current
in the ohmic regime at 310 K and is limited by experimental resolution
of preamplifier $\sim 1nV$ (not shown here). Cooling the system, BBN can be
observed along with the non-linear conductivity. A typical example is
reported in figure \ref{noise}, where the NBN value is limited by the resolution below
I$^{*}$, then increases sharply at I$^{*}$ and tends to be quiet at large
current, in agreement with successive pinning, depinning and sliding
regimes. We conclude that the electronic
transport properties of (PO$_{2}$)$_{4}$(WO$_{3}$)$_{4}$ changes
at T$\lesssim$290K, in the structurally modulated state, and both
non-linear DC conductivity and BBN appear. These peculiar electronic
properties associated to the periodic lattice distortion are strong
experimental facts for a CDW phase~\cite{monceau}. It is worth noting that
this is the first report of such a sliding collective mode in the
MPTB family.

As discussed above, (PO$_{2}$)$_{4}$(WO$_{3}$)$_{4}$
is the only member of MTBP's where W atoms are located into zigzag
chains, compared to planes for all other members with m$>$2. Then,
it can be described from its band structure~\cite{Canadell1990}
as a quasi-1D conductor. We measured here the electric field for depinning
in the range of 0.05 V/cm, which is very typical of (weakly pinned)
1D CDW and with an incommensurate modulation vector. To compare, no evidence of depinning/sliding was found in
the quasi 2D bronze Kx ($m$=4) with electric field of 0.12 V/cm~\cite{Kolincio2016}.
Our results are then consistent with the idea that a small CDW dimensionality is a key parameter for achieving small depinning thresholds. Another difference in physical properties compared to higher dimension bronzes is the absence here of a significant magnetoresistance in the CDW state,
 compared to the enhanced one observed in several others ($i.e.$, $m$=6, doped $m$=4 and $m$=10~\cite{Kolincio2016,Kolincio2020,Kolincio2016b}.

In general, quasi 1D CDW can be well described by weak electron-phonon coupling scenarios.
To better understand this observation, we have analysed the temperature variation scattered
intensity of the first order satellite peak as measured by XRD. The
idea behind is that, as far as weak displacements are assumed, this
intensity is proportional to the square of the CDW gap, providing a direct
link between electronic and structural degrees of freedom \cite{Gruner1988}.
In principle, the BCS gap equation and its deduced temperature variation
should be solved self-consistently and numerically. However, a very
good and practical analytical approximation can be used \cite{Gross1986}. As shown in the fig. \ref{fig2}, there is a very good
agreement between the experimental data and the gap interpolation formula used in the weak coupling limit.

\section{Discussion}

The member $m$=2 of the monophosphate tungsten bronze family differs structurally from the rest of the series due to its isolated zigzag chains of tungsten atoms, rather than the more typical layered W-slabs. This distinctive feature makes it the only compound in the family to exhibit a quasi-one-dimensional CDW. The difference is evident from the transition temperature, which is significantly higher (T$_\mathrm{CDW}$ = 290~K) and deviates from the trend observed across the series~\cite{Roussel2000a, Roussel2001}. The structural simplicity of this system enabled a comprehensive investigation of the charge density wave state.

We have examined the system from multiple perspectives: structurally using diffraction and MD simulations; dynamically through DS and IXS; and electronically $via$ resistivity measurements. The results converge to form a coherent picture of the phase transition and its nature.

The first set of observations concerns the pre-transitional behavior as the system evolves from the high-temperature phase to the CDW phase. DS around the low-temperature satellite peaks reveals a dynamical origin, confirmed by temperature-dependent IXS measurements. The phonon correlation length diverges near T$_\mathrm{CDW}$, displaying isotropic behavior in reciprocal space. MD simulations show highly correlated displacements of oxygen atoms surrounding the tungsten atom centers. These displacements are responsible for the low-energy phonon mode, which softens to $E = 0.80(15)$~meV just above the transition temperature (T$_\mathrm{CDW}$+10~K) at the corresponding wavevector \textbf{q}$_\mathrm{CDW}$. The softening follows the expected temperature evolution of a CDW system; however, the presence of a strong elastic signal above the transition suggests that the phonon may not fully freeze, masking the inelastic component. In a weak electron-phonon coupling regime, full phonon softening is typically observed~\cite{Pouget2016}. In contrast, in strong coupling scenarios, no apparent softening is present, and a pre-transitional elastic signal appears before T$_\mathrm{CDW}$, as seen in other systems~\cite{Pouget2024, Ilakovac2021, Subires2023}. Our results indicate that this compound may exhibit features of both regimes, highlighting the complexity of the transition.\\

The transition into the CDW phase is also evident in diffraction measurements. The structural distortion primarily involves oxygen atoms, consistent with the phonon and MD analysis. The crystal structure changes from an approximately orthorhombic symmetry ($Pmcn$) to monoclinic ($P2_1/m$). The main symmetry modes involved in this order-disorder transition are LD4 and GM1, as identified through critical wavevector analysis and MD simulations. The latter show that the yaw angle libration of the tungsten octahedra, related primarily to oxygen displacements, follows a modulation described by a phase shift of $2\pi/4$ between neighboring chains along the $b$-direction. Order-disorder transitions are typically associated with strong electron-phonon coupling. However, we note that the monoclinic distortion in the HS phase is not fully understood, which may affect our interpretation of the transition type.

Focusing now on the electronic aspects of the CDW, the tungsten atoms exhibit only a weak dimerization along the chain. This is indicative of a weak-coupling CDW, with a gap consistent with BCS theory. The temperature dependence of the satellite peak intensity supports this view. Resistivity measurements also show minimal electronic anomalies at the transition, reinforcing the idea that the structural component plays the dominant role. This may also explain the satellite wavevector \textbf{q}$_\mathrm{CDW} = 0.245\mathbf{b}^* + 0.02\mathbf{c}^*$, which does not correspond to a nesting vector of the Fermi surface. Interestingly, resistivity data reveal non-ohmic behavior emerging below the transition temperature of $\sim$290~K. This is typically attributed to CDW pinning and sliding modes and is commonly observed in quasi-1D and 1D conductors.\\

\section{Conclusion}

At first glance, this system appears to follow a classical CDW instability model, with apparent nesting. The $m$=2 member of the monophosphate tungsten bronze family undergoes a CDW transition at 290~K, characterized by an incommensurate wavevector of $0.245\mathbf{b}^* + 0.02\mathbf{c}^*$. DS around the satellite positions reveals an isotropic distribution of low-energy phonons, which condense into Bragg peaks in the CDW ground state. 

The structural transition involves tilting of both W-octahedra and P-tetrahedra, with oxygen displacements reaching up to 0.2~\AA. These rearrangements generate strong satellite intensities, but do not significantly affect the tungsten positions along the chain. This suggests a CDW of weak electron-phonon coupling character, further supported by resistivity measurements. The satellite wavevector describes correlations between adjacent W-chains, a finding corroborated by MD simulations, which reveal a fourfold periodicity along the $b$-axis.\footnote{Additional details can be found in the doctoral theses of Arianna Minelli and Elen Duverger-Nedellec~\cite{Minelli2018, Elen2017}.}\\

The combination of experimental techniques used in this study provides a comprehensive view of the CDW transition, illustrating its inherent complexity. The new phase is best described by both structural and electronic behavior -- yet, in this case, the structural reorganization appears only loosely connected to the electronic CDW. The data suggest that the lattice degrees of freedom play a central role in driving the transition.

\appendix
\section{Figures}

\begin{figure}
	\centering
		\includegraphics[scale=0.4]{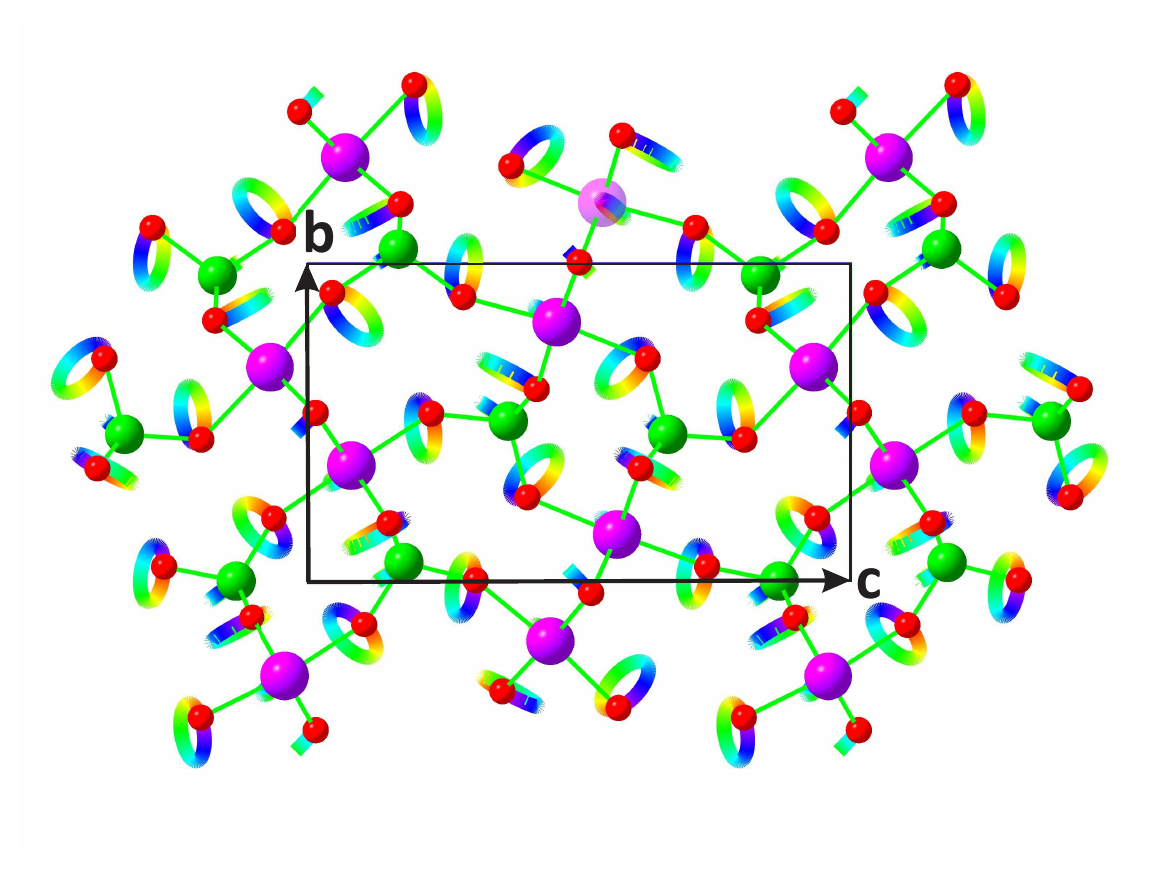}
	\caption{Projection along \textbf{a} of $P_4W_4O_{20}$ at 80K exhibiting for each atoms the Lissajou traces of the atomic displacements due to the modulation. The displacements have been magnified 2.5 times. Events occurring at the same t have the same colors in the traces. Molecoolqt [C. B. Hubschle and B. Dittrich, J. Appl. Cryst. (2011). 44, 238-240] has been used to create this figure.}
	\label{figure-4}
\end{figure}

\begin{figure}
\includegraphics[width=0.5\textwidth]{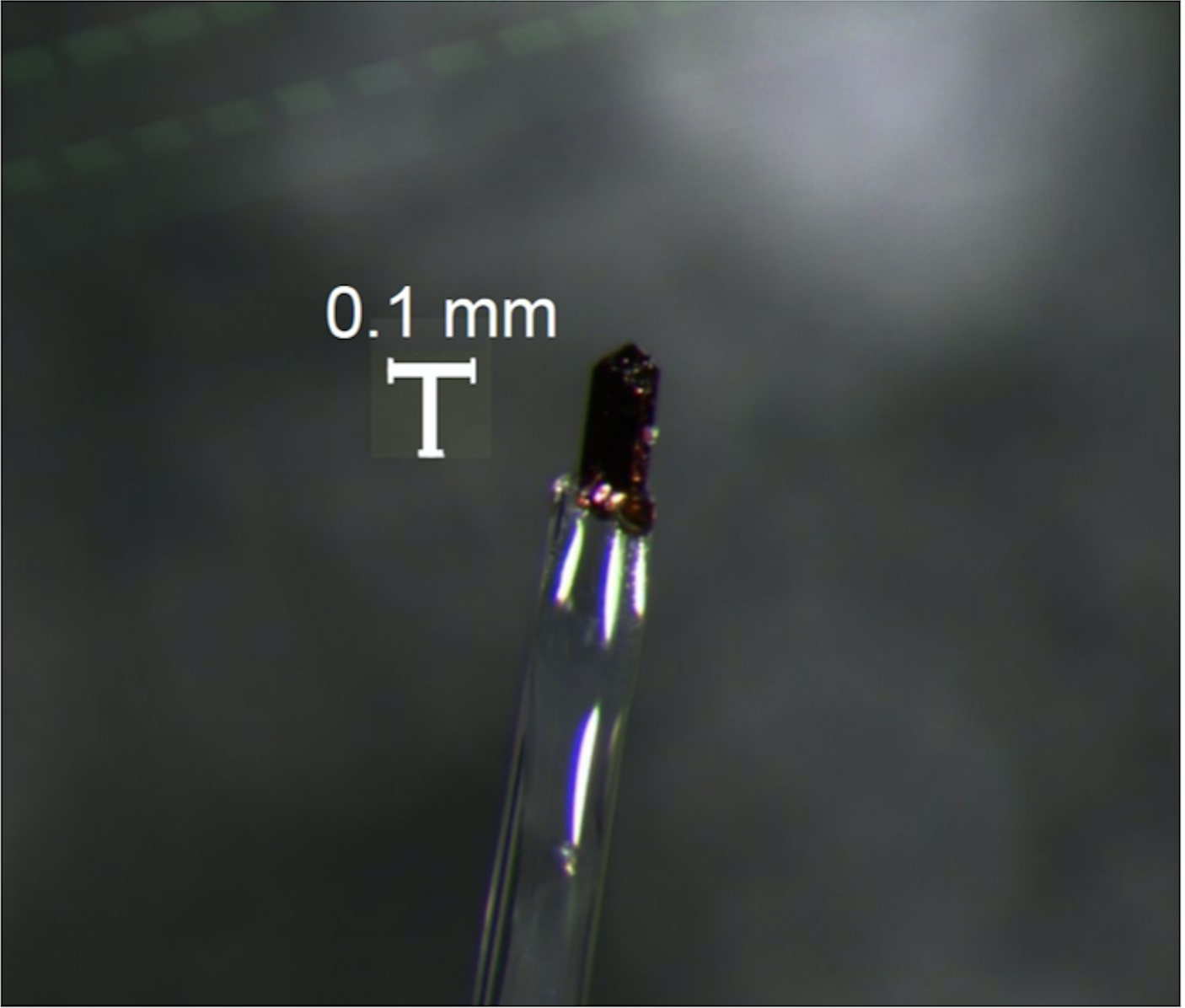}
\caption{Picture on the microscope of the \textit{m}=2 sample with a truncated parallelepipedic shape, glued on a glass capillary. }
\label{Samplem2}
\end{figure}

\begin{figure}
\includegraphics[width=0.75\textwidth]{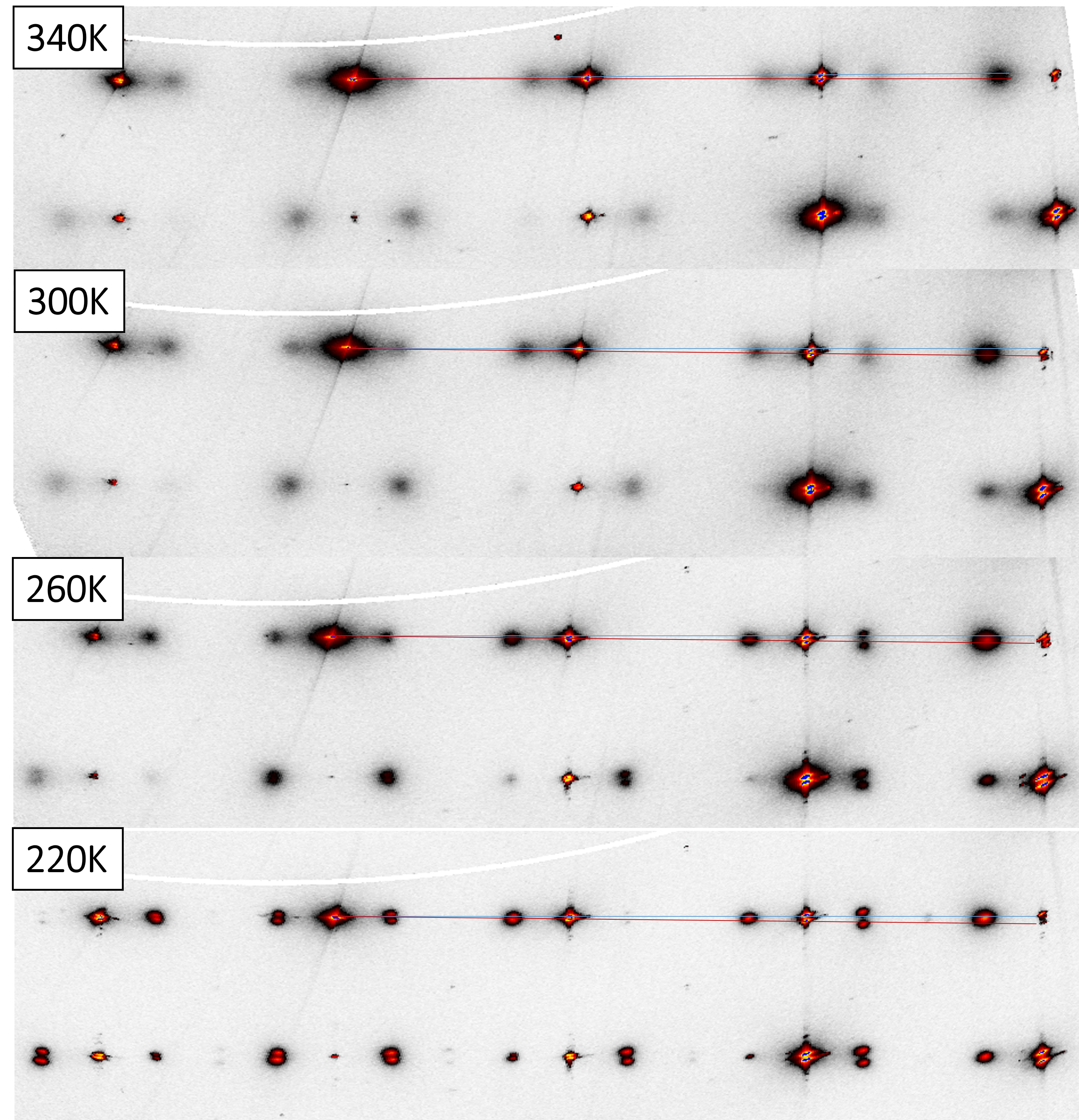}
\caption{0KL diffuse scattering maps above the CDW transition temperature. The red and blue lines follow the two domains showing the monoclinic distortion. The temperatures are referred to the cryostream temperatures, while the sample is probably warmer (~100K).}
\label{HighQDS}
\end{figure}

Here are reported few pictures. In figure \ref{figure-4}, the Lissajou traces of the atomic displacements due to the modulation in the low-temperature phase of $m$=2 taken by the in house diffraction. In figure \ref{Samplem2}, the sample used for DS and IXS measurement at ID28 (ESRF) . The figure \ref{HighQDS} shows the high-Q region measured in ID28 that permits to identify the monoclinic distortion. The temperatures are referred to the cryostream, not referred to the temperature of the sample. It is probable that the beam was heating the sample $\sim$100K, thus the temperature of transition will be $\sim$190K in this configuration.

\begin{figure}[]
\includegraphics[width=1\textwidth]{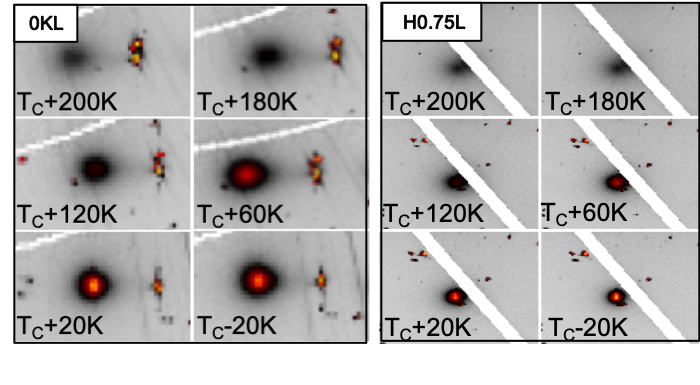}
\caption{0KL and HK0.75 maps at different temperatures, around the satellite position at 0,0.75,$\overline{5}$.}
\label{Maps-diffuse}
\end{figure}

The figure \ref{Maps-diffuse} shown the zoom of the 0KL and HK0.75 maps around the satellite 0,0.75,$\overline{5}$. We can see the dependence of the diffuse to the temperature and the strong intensity of the new satellite reflection at the CDW ground state.
\\

\begin{figure}
\includegraphics[width=\textwidth]{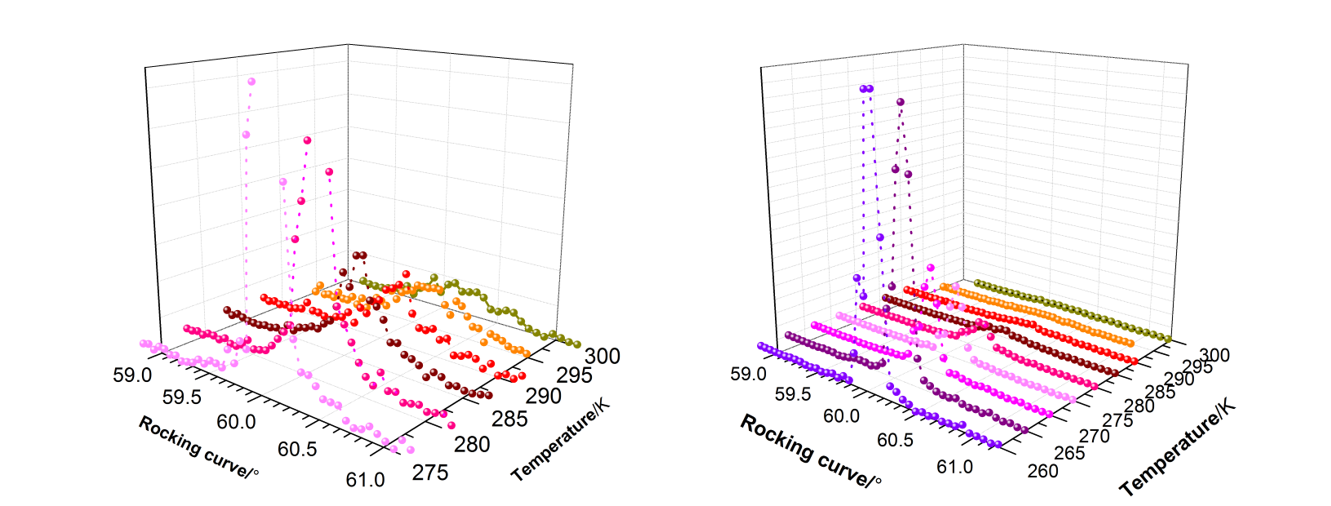}
\caption{Temperature dependence on the rocking curve for the satellite of \textit{m}=2 made by the spectrometer on the CDW satellite, \textbf{q}=$\sim$-0.25\textbf{b*}, in the point Q(0 -0.75 6), from 300 K to 275 K, on the left, and until the transition, on the right, where the intensity  increase and the Bragg peak is formed.}
\label{Rockingcurve}
\end{figure}

\begin{figure}
\includegraphics[width=\textwidth]{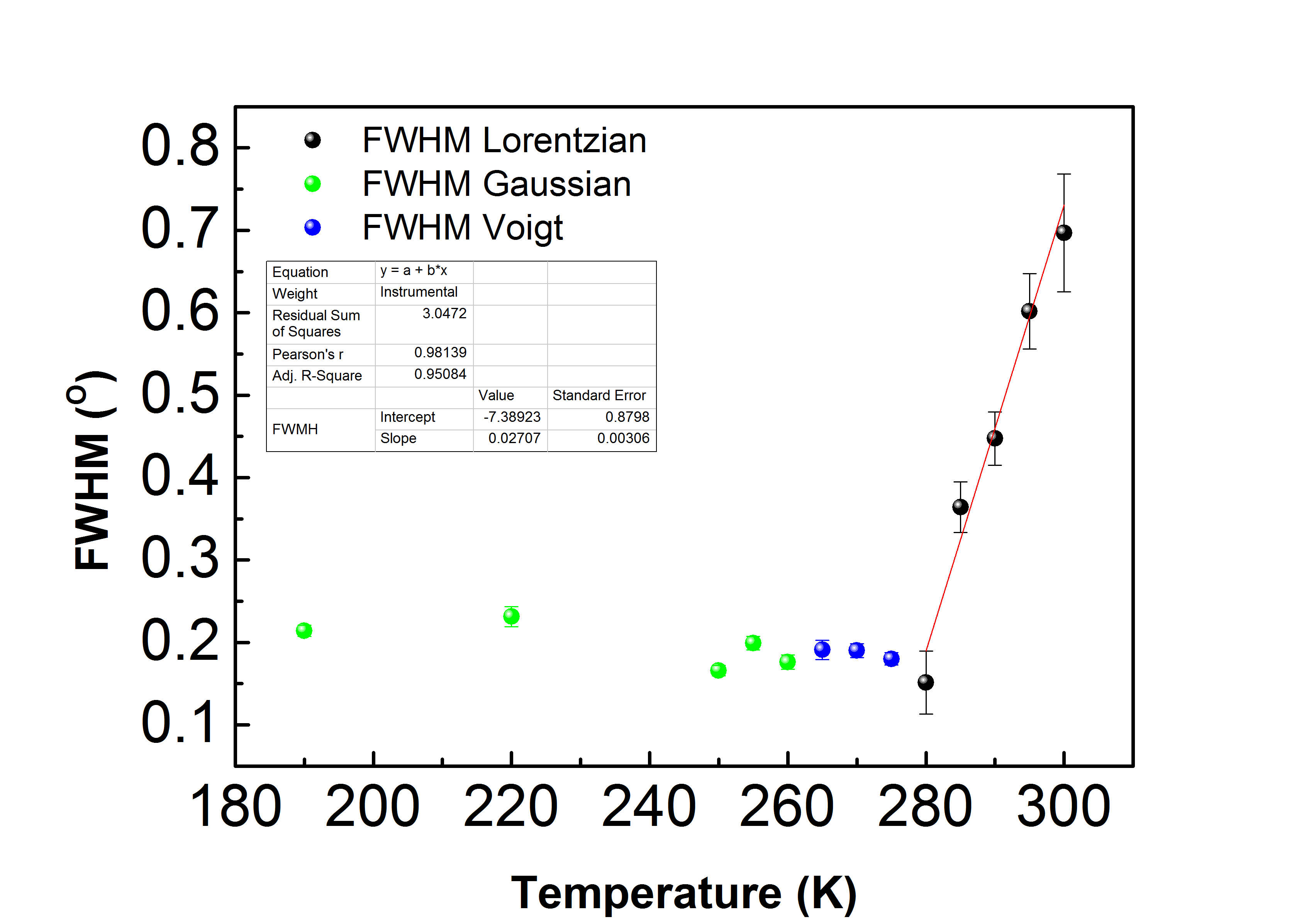}
\caption{Full Width Half Maxima (FWHM) of the Pseudo-Voigt, Gaussian and Lorentzian fitting are reported for the IXS scans at different temperatures.}
\label{FWHM}
\end{figure}

For the IXS measurement, the temperature of transition was checked through an elastic measurement of the rocking curve of the Bragg peak 0,0.75,$\overline{6}$, figure \ref{Rockingcurve} and its fitting was rather complicated. The transition should be reached, when the FWHM is just the resolution, so the order parameter converge to infinite. This temperature is obtained with the crossing of the linear fitting of the Lorentzian FWHM at y=0.2, which is more or less the resolution of the instrument (or the broadening of this peak in particular), which gives 280K, as shown in figure \ref{FWHM}.

\begin{figure}
	\centering
		\includegraphics[width=\textwidth]{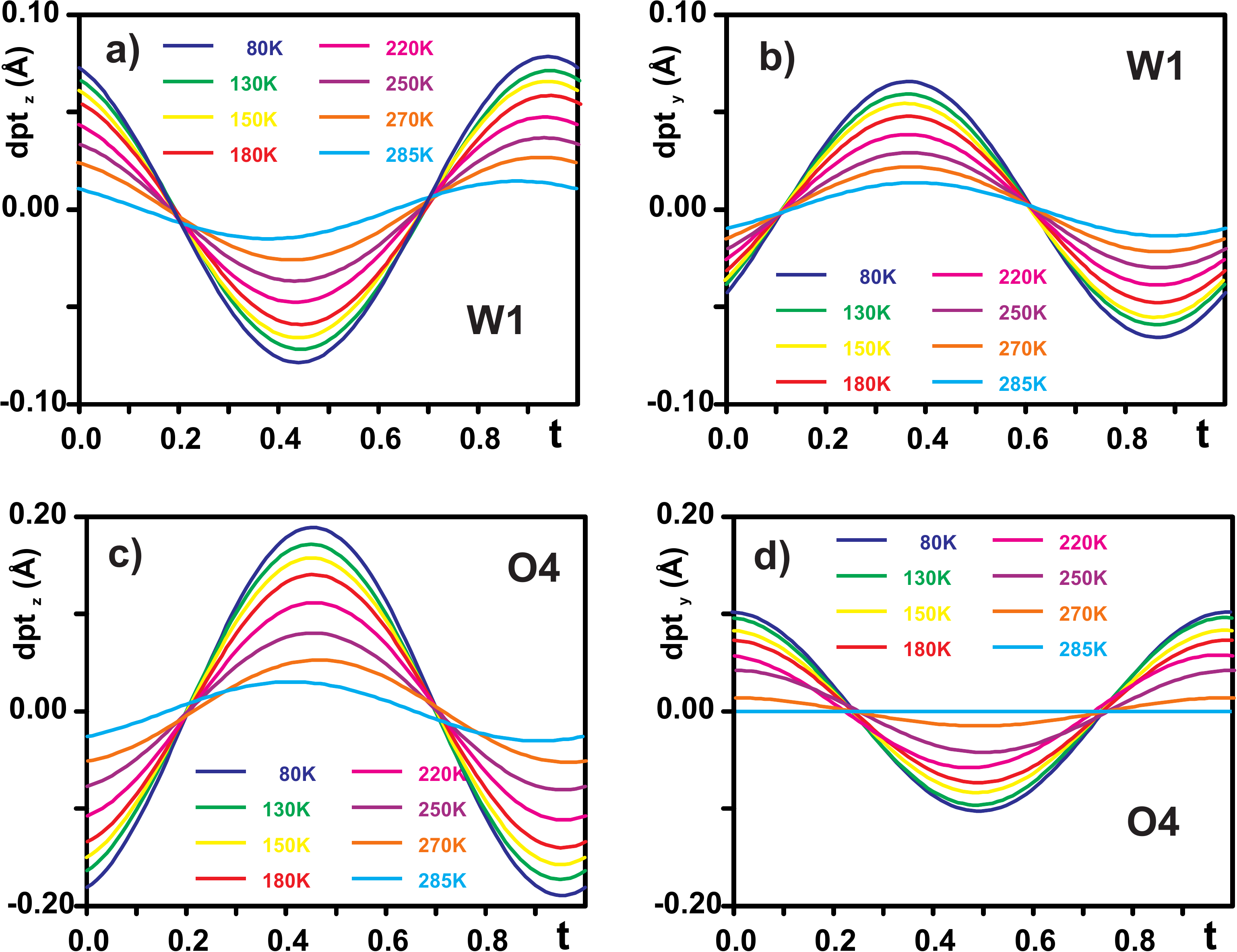}
	\caption{Evolution versus t of the atomic displacements for W1 and O4 for different temperatures.}
	\label{figure-SI1}
\end{figure}

\section{Tables}

\begin{table}

\begin{tabular}{|l|l|l|}

\hline
   & RT & 80K \\
\hline
chemical formula &   $P_4W_4O_{20}$     &   $P_4W_4O_{20}$     \\
Space group &    Pmcn    &           P2$_1$/m  \\ 
 \textbf{a} (\AA)              &   5.22786(5)     &     5.2302(2)    \\
  \textbf{b} (\AA)              &    6.55814(11)    &   6.5427(5)     \\
  \textbf{c} (\AA)              &    11.1883(2)    &   11.1823(4)     \\
  $\alpha$ ($^{\circ}$)              &  90.0583(15)      &  89.900(5)      \\
  $\beta$ ($^{\circ}$)              &  90.0118(12)      &   89.984(3)      \\
  $\gamma$ ($^{\circ}$)              &  90.0024(10)      &  89.979(5)      \\
Volume (\AA$^3$) &  383.591(12)      &  382.65(4)      \\
q wave vector &   -     &    0.2502(11) $ \vec{b}^\star $    \\
Z				&    1    &        \\				
		h limit	    & $-8 \leq h \leq  8 $&   $-8 \leq h \leq  8 $     \\ 					
    k limit			& $-11 \leq k \leq 10$ &  $-11 \leq k \leq 11$      \\					
    l limit     & $-18 \leq l \leq 18$ &   $-17 \leq l \leq 18$     \\
reflections $\left[I\geq 3 \sigma (I)\right]$ 	&   952     &   2254     \\
\qquad main &   952     &    911    \\
\qquad 1$^{st}$ order satellite &  -      &   1343     \\

Rint ($\%$) &     0.05   &        \\
no of refinement parameters &   41     &   48     \\
R$_F$ ($\%$) &    2.08    &   3.49     \\
\qquad R$_{F,0}$ ($\%$) &  -      &  2.45      \\
\qquad R$_{F,1}$ ($\%$) &  -      &   6.23     \\

\hline
\end{tabular}	\\
	\caption{Refinement parameters}
	\label{Append1}
\end{table}

\begin{table}
\begin{tabular}{|l|l|l|l|l|l|}
\hline
atom & & x & y & z & u$_{iso}$ \\
\hline
W1 & 1 &  0.25 & -0.16856(2) &  0.438765(13) &  0.00569(6) \\
P1 & 1 &  0.25 &  0.47108(16) &  0.65538(10) &  0.0067(2)\\
O1 & 1 &  0.25 &  0.6695(5) &  0.5859(4) &  0.0248(14)\\
O2 & 1 &  0.25 &  0.5209(6) &  0.7871(3) &  0.0188(10)\\
O3 & 1 &  0.0124(9) &  0.3535(7) &  0.6263(4) &  0.0560(15)\\
O4 & 1 &  0 &  1 &  0.5 &  0.0283(12)\\
\hline
\end{tabular}	\\
	\caption{Positional parameters at RT}
	\label{Append2}
\end{table}

\begin{table}
\begin{tabular}{|l|l|l|l}
\hline
atom  & $u_{11}$ & $u_{22}$ & $u_{33}$  \\
\hline 
W1 &  0.00406(11) &  0.00787(11) &  0.00514(11)  \\
P1 &  0.0073(4) &  0.0081(4) &  0.0046(4)  \\
O1 &  0.043(3) &  0.0158(19) &  0.0150(18)  \\
O2 &  0.029(2) &  0.0235(17) &  0.0041(14)  \\
O3 &  0.069(3) &  0.077(3) &  0.0227(19)  \\
O4 &  0.031(2) &  0.044(3) &  0.0098(16)  \\
\hline
atom  & $u_{12}$ & $u_{13}$ &$u_{23}$ \\
\hline 
W1 &    0 &  0 & -0.00048(4) \\
P1 &    0 &  0 & -0.0002(3) \\
O1 &    0 &  0 &  0.0100(12) \\
O2 &    0 &  0 & -0.0051(11) \\
O3 &   -0.065(2) & -0.0160(18) &  0.0060(16) \\
O4 &   0.0271(19) &  0.0098(15) &  0.0107(15) \\
\hline

\end{tabular}	\\
	\caption{ADP harmonic parameters at RT}
	\label{adp-par-rt}
\end{table}

\begin{table}
\begin{tabular}{|l|l|l|l|l|l|l|}
\hline
atom & & occ & $\left\langle x \right\rangle$ & $\left\langle y \right\rangle$ & $\left\langle z \right\rangle$ & $u_{iso}$ (\AA$^2$) \\
\hline
W1 & 1 &  &  0.25 & -0.16828(3) &  0.438710(15) &  0.00414(7) \\
   &  & s,1 &  0 &  0.0091 & -0.004220(18) &   \\
   &  & c,1 &  0 & -0.0043 &  0.005616(19) &   \\
P1 & 1 &  &  0.25 &  0.47061(19) &  0.65512(11) &  0.0055(3) \\
   &  & s,1 &  0 &  0.0092 & -0.00727(12) &  \\
   &  & c,1 &  0 &  0.0046 & -0.00442(12) &  \\
O1 & 1 &  &  0.25 &  0.5213(5) &  0.7872(3) &  0.0124(10) \\
   &  & s,1 &  0 &  0.0066 & -0.0073(3) &  \\
   &  & c,1 &  0 & -0.0296 &  0.0005(4) &  \\
O2 & 1 &  &  0.25 &  0.6673(6) &  0.5837(4) &  0.0192(13) \\
   &  & s,1 &  0 &  0.0153 & -0.0035(4) &  \\
   &  & c,1 &  0 &  0.0205 &  0.0141(4) &  \\
O3 & 1 &  &  0 &  1 &  0.5 &  0.0287(15) \\
   &  & s,1 &  0.0003(9) &  0.0129(8) & -0.0072(4) &  \\
   &  & c,1 &  0 &  0 &  0 &  \\
O4 & 1 &  &  0.0116(10) &  0.3515(7) &  0.6270(3) &  0.0536(17) \\
   &  & s,1 & -0.0003(9) &  0.0078(7) & -0.0043(3) &  \\
   &  & c,1 &  0.0004(9) &  0.0136(8) & -0.0163(4) &  \\
\hline
\end{tabular}	\\
	\caption{Positional parameters at 80K}
	\label{pos-par80k}
\end{table}

\begin{table}
\begin{tabular}{|l|l|l|l}

\hline 
atom  & $u_{11}$ & $u_{22}$ & $u_{33}$ \\
\hline 
W1 &    0.00276(11) &  0.00724(11) &  0.00242(11) \\
P1 &    0.0064(5) &  0.0075(5) &  0.0026(5)  \\
O1 &    0.0215(19) &  0.0138(15) &  0.0018(14)  \\
O2 &    0.039(3) &  0.0126(17) &  0.0064(16)\\
O3 &   0.030(3) &  0.049(3) &  0.0070(18)  \\
O4 &    0.072(3) &  0.075(3) &  0.0138(17)  \\
\hline
atom  & $u_{12}$ & $u_{13}$ &$u_{23}$ \\
\hline 
W1 &      0 &  0 & -0.00009 \\
P1 &      0 &  0 & -0.000393 \\
O1 &      0 &  0 & -0.001928 \\
O2 &      0 &  0 &  0.003109 \\
O3 &     0.029(2) &  0.0089(18) &  0.0094(19) \\
O4 &    -0.067(3) & -0.0148(18) &  0.0086(18) \\
\hline

\end{tabular}	\\
	\caption{ADP harmonic parameters at 80K}
	\label{adp-par80k}
\end{table}

\clearpage

\begin{table}[]
\centering
\setlength{\tabcolsep}{6pt}
\renewcommand{\arraystretch}{1.15}

\begin{tabularx}{\columnwidth}{|l|X|X|}
\hline
Polyhedron & Distances (\AA) & Angles ($^{\circ}$) \\
\hline

$WO_6$
&
$d_{W1-O1^{(1)}} = 1.955(5)\pm0.011$ \newline
$d_{W1-O2^{(2)}} = 1.952(5)\pm0.011$ \newline
$d_{W1-O3^{(2)}} = 1.843(3)\pm0.003$ \newline
$d_{W1-O3^{(3)}} = 1.843(3)\pm0.003$ \newline
$d_{W1-O4^{(3)}} = 1.964(7)\pm0.006$ \newline
$d_{W1-O4^{(4)}} = 1.964(7)\pm0.006$
&
$\widehat{O1^{(1)}\!-\!W1\!-\!O2^{(2)}} = 176.00(13)\pm1.00$ \newline
$\widehat{O1^{(1)}\!-\!W1\!-\!O3^{(2)}} = 91.66(15)\pm0.93$ \newline
$\widehat{O1^{(1)}\!-\!W1\!-\!O3^{(3)}} = 91.66(15)\pm0.93$ \newline
$\widehat{O1^{(1)}\!-\!W1\!-\!O4^{(5)}} = 88.62(17)\pm0.52$ \newline
$\widehat{O1^{(1)}\!-\!W1\!-\!O4^{(4)}} = 88.62(17)\pm0.52$ \newline
$\widehat{O2^{(2)}\!-\!W1\!-\!O3^{(2)}} = 91.18(16)\pm0.74$ \newline
$\widehat{O2^{(2)}\!-\!W1\!-\!O3^{(3)}} = 91.18(16)\pm0.74$ \newline
$\widehat{O2^{(2)}\!-\!W1\!-\!O4^{(5)}} = 88.48(18)\pm0.74$ \newline
$\widehat{O2^{(2)}\!-\!W1\!-\!O4^{(4)}} = 88.48(18)\pm0.74$ \newline
$\widehat{O3^{(2)}\!-\!W1\!-\!O3^{(3)}} = 90.43(18)\pm0.23$ \newline
$\widehat{O3^{(2)}\!-\!W1\!-\!O4^{(5)}} = 178.60(2)\pm0.40$ \newline
$\widehat{O3^{(2)}\!-\!W1\!-\!O4^{(4)}} = 90.7(2)\pm0.3$ \newline
$\widehat{O3^{(3)}\!-\!W1\!-\!O4^{(5)}} = 90.7(2)\pm0.3$ \newline
$\widehat{O3^{(3)}\!-\!W1\!-\!O4^{(4)}} = 178.60(2)\pm0.40$ \newline
$\widehat{O4^{(5)}\!-\!W1\!-\!O4^{(4)}} = 88.3(3)\pm0.4$
\\
\hline

$PO_4$
&
$d_{P1-O1^{(0)}} = 1.522(6)\pm0.009$ \newline
$d_{P1-O2^{(0)}} = 1.527(3)\pm0.016$ \newline
$d_{P1-O4^{(0)}} = 1.506(7)\pm0.005$ \newline
$d_{P1-O4^{(6)}} = 1.506(7)\pm0.005$
&
$\widehat{O1^{(0)}\!-\!P1\!-\!O2^{(0)}} = 109.2(2)\pm0.3$ \newline
$\widehat{O1^{(0)}\!-\!P1\!-\!O4^{(0)}} = 108.6(2)\pm0.3$ \newline
$\widehat{O1^{(0)}\!-\!P1\!-\!O4^{(7)}} = 108.6(2)\pm0.3$ \newline
$\widehat{O2^{(0)}\!-\!P1\!-\!O4^{(0)}} = 109.3(2)\pm0.3$ \newline
$\widehat{O2^{(0)}\!-\!P1\!-\!O4^{(7)}} = 109.3(2)\pm0.3$ \newline
$\widehat{O4^{(0)}\!-\!P1\!-\!O4^{(7)}} = 111.9(4)\pm0.7$
\\
\hline
\end{tabularx}

\caption{Quantification of polyhedra distortion due to the modulation at 80 K (average value $\pm$ deviation due to the modulation). Symmetry codes: 
(0) $x,y,z$; (1) $-x+\tfrac12,-y+\tfrac12,z-\tfrac12$; 
(2) $x,y-1,z$; (3) $x+\tfrac12,-y+1,-z+1$; 
(4) $-x,-y,-z+1$; (5) $x+\tfrac12,-y,-z+1$; 
(6) $x+\tfrac12,y,z$; (7) $-x+\tfrac12,y,z$.}
\label{poly-dis}

\end{table}



\ack{Acknowledgements}

The authors would like to thank Denis Gambetti for technical support during the experiment in ID28. In this work the authors acknowledge the computing support of the Slovak Academy of Science. Additionally, this work was partially supported by the found for young researchers of the CMAC network. A. M. ackowledges financial support from ``COMPLEXORDER'' ERC No. 788144 and the work of A.M. has been supported by the DOE Office of Science.





\end{document}